\documentclass[12pt,a4paper]{article}
\usepackage[top=3cm, bottom=2.5cm, right=2.5cm, left=2.5cm]{geometry}
\usepackage[english]{babel}
\usepackage[cp1250]{inputenc}
\usepackage[headings]{fullpage}
\usepackage[intlimits]{amsmath}
\usepackage{amssymb}
\usepackage{graphicx}
\usepackage{xcolor}
\usepackage{mathrsfs}
\usepackage{psfrag}
\usepackage{eprint}
\usepackage{hyperref}

\DeclareMathOperator{\sgn}{sgn}
\newcommand{\sig}{\sigma}
\newcommand{\kap}{\kappa}
\newcommand{\beq}{\begin{equation}}
\newcommand{\beql}[1]{\begin{equation}\label{#1}}
\newcommand{\eeq}{\end{equation}}
\newcommand{\bea}{\begin{eqnarray}}
\newcommand{\eea}{\end{eqnarray}}
\newcommand{\lamd}{\dot{\lambda}}
\newcommand{\lamp}{\lambda_p}
\newcommand{\lamg}{\lambda_g}
\newcommand{\kapd}{\dot{\kappa}}
\newcommand{\kappat}{\kappa_n}
\newcommand{\kapt}{\kappa_n}
\newcommand{\kaph}{\hat\kappa}
\newcommand{\kaptb}{\bar{\kappa}_n}
\newcommand{\kapb}{\bar{\kappa}}
\newcommand{\eps}{\varepsilon}
\newcommand{\epse}{\varepsilon_e}
\newcommand{\epsp}{\varepsilon_p}
\newcommand{\epsdp}{\dot\varepsilon_p}
\newcommand{\sigY}{\sigma_Y}
\newcommand{\sigc}{\sigma_c}
\renewcommand{\lg}{l_g} 
\newcommand{\half}{\mbox{$\frac{1}{2}$}}
\newcommand{\Lp}{L_p}
\newcommand{\Ip}{I_p}
\newcommand{\Ie}{I_e}
\newcommand{\cotan}{\mathrm{cotan}}
\newcommand{\dx}{\mathrm{d}x}
\newcommand{\dxi}{\mathrm{d}\xi}

\begin{document}

\title{Softening Gradient Plasticity: Analytical Study of Localization under Nonuniform Stress}
\author{Milan Jir\'{a}sek, Jan Zeman and Jaroslav Vond\v{r}ejc\\
Department of Mechanics, Faculty of Civil Engineering\\
Czech Technical University in Prague, Czech Republic}

\maketitle

\begin{abstract}
Localization of plastic strain induced by softening can be objectively 
described by a regularized plasticity model that postulates a dependence
of the current yield stress on a nonlocal softening variable defined by
a differential (gradient) expression.
This paper presents analytical solutions of the one-dimensional localization
problem under certain special nonuniform stress distributions. 
The one-dimensional problem can be interpreted as describing either
a tensile bar with variable cross section, or a beam subjected to
a nonuniform bending moment. Explicit as well as implicit gradient 
formulations  are considered. The evolution of the plastic strain profile
and the shape of the load-displacement diagram are investigated.
It is shown that even if the local constitutive law exhibits softening
right from the onset of yielding, the global load-displacement diagram
has a hardening part. The interplay between the internal length scales
characterizing the material and the geometry is discussed.
\end{abstract}

\section{Introduction}

\subsection{Gradient plasticity as a localization limiter}

Stress-strain diagrams of quasibrittle materials typically exhibit
softening, which is caused by propagation and coalescence of defects
in the microstructure,
and at the macroscopic scale is manifested by decreasing stress at
increasing strain. Softening can naturally be incorporated into damage
models but, alternatively, can be modeled within the framework of
plasticity with yield limit degradation. 

In the context of the standard continuum approach, softening may lead
to localization of plastic strain into an arbitrarily small volume and,
consequently, to the pathological sensitivity of the numerical results
obtained by the finite element method to the size of elements used in the
analysis. Objective, mesh-insensitive description of the localization
phenomenon requires an enhancement of the governing equations by non-standard
terms that act as localization limiters and regularize the solutions.
One popular class of localization limiters is based on the incorporation
of second or higher gradients of the softening variable into the softening law
\cite{BazBelCha84}.
Many gradient plasticity formulations have been inspired by the pioneering
work of Aifantis and coworkers \cite{Aifantis84}, others by the implicit
gradient approach used in damage mechanics \cite{Peerlings96}. 

Localization properties of gradient plasticity models have often been
studied by investigating the bifurcation from a uniform state
in the idealized one-dimensional setting; see \cite{JirRol09b} for a summary.
The aim of this paper is to extend the localization analysis to cases
with a non-uniform stress field, which arises for instance in a bar under
uniaxial tension due to a variation of the sectional area. To make the 
problem amenable to an analytical solution, we restrict our interest
to static equilibrium under vanishing body forces, using the 
small-strain assumptions.  

\subsection{One-dimensional softening plasticity model}
\label{subsec:1.2}

In the one-dimensional setting, the standard elastoplastic model is based
on the additive split of the total strain $\eps$ into the elastic part
$\epse$ and plastic part $\epsp$, with stress $\sig$ linked to the elastic
strain by Hooke's law
\beql{eqgp1}
\sig = E\epse = E(\eps-\epsp)
\eeq
The yield function that identifies elastic, plastic and inadmissible states 
can be defined as\footnote{The fact that we assume 
a symmetric response in tension and in compression, 
and that the hardening or softening is taken as isotropic, does not impose
any restrictions on generality, since we will investigate localization under 
monotonic loading and the sign of stress or of the plastic strain increment
will never change.} 
\beql{eqgp2}
f(\sig,\kappa) = \vert \sig \vert - \sigY(\kappa)
\eeq
where $\sigY$ is the current yield stress, evolving as a function of an internal
variable $\kappa$, to be specified later.
The hardening-softening law will be considered in the simplest linear form,
\beql{eqgp2a}
\sigY(\kappa)=\sig_0+H\kappa
\eeq
where $\sig_0$ is the initial yield stress and $H$ is the plastic modulus.
Positive values of $H$ correspond to hardening and negative values to softening.
Our attention will be focused on the latter case, in which $H$ is referred to
as the softening modulus.
 
The evolution of plastic strain is formally described by the flow rule
\beql{eqgp3}
\epsdp=\lamd\sgn\sig
\eeq
and the loading-unloading conditions
\beql{eqgp4}
\lamd\ge 0, \hskip 10mm f(\sig,\kappa)\le 0, \hskip 10 mm \lamd f(\sig,\kappa)=0
\eeq
where $\lambda$ is the plastic multiplier and the dot over a symbol denotes differentiation 
with respect to time. The internal variable $\kappa$ is usually taken
as the cumulative plastic strain and defined by the rate equation
\beql{eqgp5}
\kapd=\vert\epsdp\vert
\eeq
If we consider only tensile loading (with possible elastic unloading,
but never with a reversal of plastic flow), then the plastic strain $\epsp$, cumulative
plastic strain $\kappa$ and plastic multiplier $\lambda$ are all equal.
We will use $\kappa$ as the primary symbol for (cumulative) plastic strain 
and rewrite equations (\ref{eqgp1}) and  (\ref{eqgp4}) as
\beql{eqgp6}
\sig = E(\eps-\kappa)
\eeq
\beql{eqgp7x}
\kapd\ge 0, \hskip 10mm f(\sig,\kappa)\le 0, \hskip 10 mm \kapd f(\sig,\kappa)=0
\eeq

Formally the same framework can be used for the description of an elastoplastic
moment-curvature relation that characterizes the inelastic flexural response of a beam.  
Stress and strain are then replaced by bending moment and curvature, Young's modulus $E$
by the sectional bending stiffness $EI$ (where $I$ is the sectional moment of inertia),
the initial yield stress $\sig_0$ by the initial plastic moment $M_0$, and the softening
modulus $H$ by a constant $C$ that represents the derivative of the plastic moment
with respect to the plastic part of curvature (in the post-yield range). Description
of the moment-curvature relation by a bilinear diagram is certainly a rough approximation,
but it can reflect the main features of inelastic bending and serve as a prototype model,
for which analytical solutions exist.

\subsection{One-dimensional localization problem}

It is well known that, in the one-dimensional setting, softening immediately leads to
localization of inelastic strain. If we consider a straight bar with perfectly uniform
properties subjected to uniaxial tension (induced by applied displacement at one bar end), the response
remains uniform in the elastic range and also during plastic yielding with a positive plastic
modulus. For a negative (or vanishing) plastic modulus, uniqueness of the solution is 
lost right at the onset
of softening (or of perfectly plastic yielding). 
Stress distribution along the bar must still remain uniform due to the static
equilibrium conditions (in the absence of body forces), but a given stress level can
be attained by softening with increasing plastic strain, or by elastic unloading with no
plastic strain evolution. Which cross sections unload
and which exhibit softening remains completely undetermined, 
and there is no lower bound on the total length of the
softening region(s). Therefore, infinitely many solutions exist, including solutions
with plastic strain evolution localized into extremely small regions. Since the dissipation
per unit volume is fixed and the volume of the softening material is arbitrarily small, 
failure of the bar can occur at arbitrarily small dissipation.
This theoretical deficiency of the standard softening model is at the origin of numerical
problems with pathological sensitivity of finite element solutions to the size of the elements. 
Even if the nonuniqueness of the solution is removed by a slight perturbation of the perfect
uniformity of the bar, the problem with localization of softening into arbitrarily small
regions (in fact into the weakest cross section) still persists.

Remedy is sought either in adjustment of the softening modulus proportionally to the element size
\cite{Pietruszczak81,BazOh83},
or in advanced regularization techniques that serve as localization limiters,
i.e., prevent localization of plastic strain into arbitrarily small regions and enforce
a certain minimum size of the plastic zone. Of course this minimum size is not directly prescribed, 
it is rather the outcome of the solution of the governing equations, which are
enhanced by nonstandard terms that contain (sometimes in a hidden form) at least one
new model parameter with the dimension of length. Such spatial scale
information is not contained in standard
constitutive equations based on traditional continuum mechanics. For instance, all
the material parameters used by the simple elastoplastic model in 
subsection \ref{subsec:1.2}, i.e.\ $E$, $\sigma_0$ and $H$, have the dimension of stress and 
cannot be combined into a parameter with the dimension of length. This is generally
true of all the ``local'' and ``standard-order'' continuum theories.\footnote{A characteristic
length is present in fracture mechanics, but that is not a ``pure'' continuum theory, since
it admits discontinuities in the displacement field.} 
Roughly speaking, by ``local''
we mean that the constitutive response at each material point depends on the state variables at that point only
(and possibly on their previous evolution), and by ``standard-order'' we mean that
(i) the state of deformation at each material point is fully described by the first gradient of the displacement
field while the higher gradients have no influence, (ii) no additional kinematic variables are
introduced, and (iii) gradients of internal variables are not incorporated in the constitutive equations. 

Enrichment of the constitutive model by nonstandard terms that contain a length parameter
can be achieved either by nonlocal approaches (in the narrow sense), which consider the influence
of a finite neighborhood of each material point, or by higher-order approaches, which consider
the influence of higher (than usual) gradients of the displacement or of internal variables,
or introduce additional variables characterizing the deformation state (e.g.\ micro-rotations
that are not equal to the macro-rotations derived from the displacement field), or deal with
gradients of internal variables. Approaches based on higher-order gradients are sometimes
considered as nonlocal in the broad sense, and certain specific formulations of this kind 
are even equivalent to nonlocal approaches in the narrow sense (which are also called integral-type
nonlocal approaches because they use weighted spatial averaging described by integral operators).

This study focuses on softening plasticity models enhanced by gradients of internal variables,
which means, for the simple one-dimensional model considered here, models with spatial derivatives
of the cumulative plastic strain $\kappa$. Localization properties of a wide range of such models have been   
analyzed and compared in \cite{JirRol09b}, but only for the highly idealized case of a perfectly uniform stress field.
It has been shown that some formulations suffer by serious deficiencies and thus do not need to be
considered in more detailed studies. For this reason, we will restrict our attention to two widely popular
families of models, referred to as explicit and implicit. Explicit gradient models have their origins
in the pioneering work of Aifantis \cite{Aifantis84}, while implicit gradient models found inspiration in the implicit
gradient damage model \cite{Peerlings96} and for plasticity were first proposed by Geers and coworkers 
\cite{GeeEngUba01,EngGeeBaa02,Gee04}.
In each family, we will investigate the basic version and one of its modifications:
\begin{enumerate}
\item
Explicit gradient model 
\begin{enumerate}
\item with second gradient of cumulative plastic strain,
\item with fourth gradient of cumulative plastic strain.
\end{enumerate}
\item
Implicit gradient model 
\begin{enumerate}
\item with boundary conditions at the physical boundary,
\item with ``boundary'' conditions at the boundary of the plastic zone. 
\end{enumerate}
\end{enumerate}

In contrast to \cite{JirRol09b}, we will consider nonuniform stress distribution, which can be caused
by variations of the sectional area in the problem of bar under uniaxial tension, but at the same time
can reflect the nonuniform distribution of bending moments in a beam. 
From this point of view, two basic cases can be distinguished:
\begin{enumerate}
\item
Smooth variation of stress, approximated in the vicinity of the global maximum by a concave quadratic function.
\item
Continuous variation of stress with discontinuous spatial derivative at the global maximum,
approximated by a concave piecewise linear function.
\end{enumerate}
The least regular case of a  stress field with discontinuities (due to jumps in the sectional area) is more intricate and will be addressed in a separate publication.

\begin{figure}%
\centering
\begin{tabular}{cc}
(a) & (b) \\[3mm]
\includegraphics[scale=0.4]{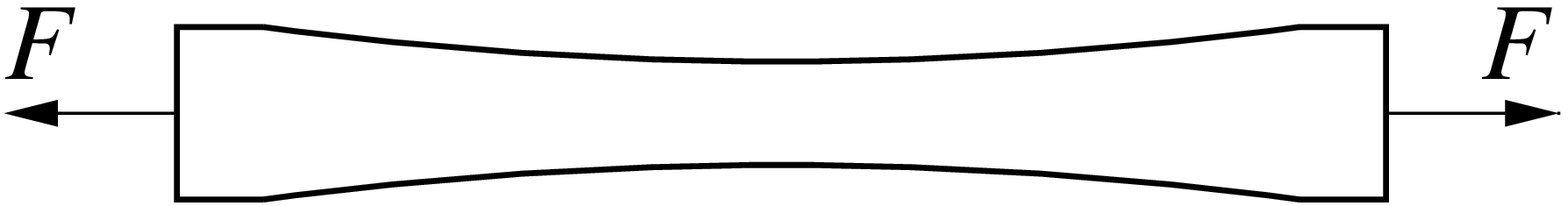}
&
\includegraphics[scale=0.4]{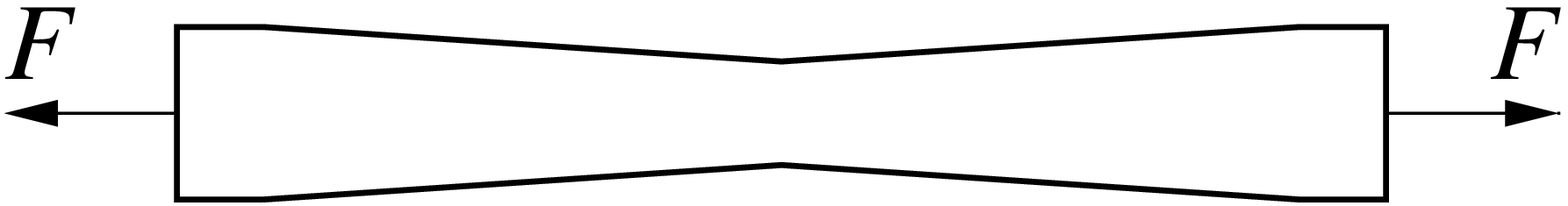}
\end{tabular}
\caption{Tensile bars with (a) smooth distribution of sectional area, 
(b) continuous but non-smooth distribution of sectional area}%
\label{fgp1}%
\end{figure}

\begin{figure}%
\centering
\begin{tabular}{ccc}
(a) & (b) \\[3mm]
\includegraphics[scale=0.7]{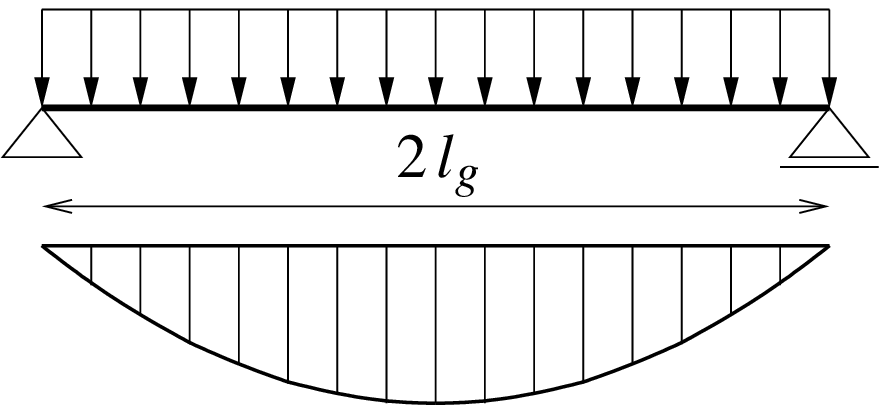}
&
\includegraphics[scale=0.7]{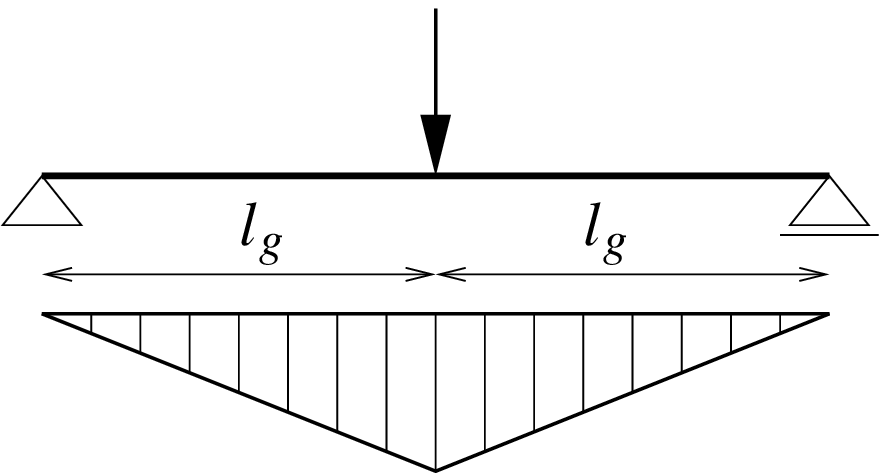}
\end{tabular}
\caption{Simply supported beam with (a) quadratic distribution of bending moments, (b) linear distribution of bending moments}%
\label{fbeam}%
\end{figure}

The first case corresponds to  an axially stretched bar of a dog-bone shape;
see Fig.~\ref{fgp1}a. The  stress distribution that allows for an analytical solution is given by
the quadratic function
\beql{eqgp7}
\sig(x) = \sig(0) + \half\sig''(0)x^2 = \sigc \left(1-\frac{x^2}{\lg^2}\right)
\eeq
where the origin of the spatial coordinate $x$ is placed at the weakest section with maximum stress $\sigc=\sig(0)$,
and $\lg=\sqrt{-2\sig(0)/\sig''(0)}$ is a parameter that sets the length scale of stress variation.
Expression (\ref{eqgp7}) can be considered as the Taylor expansion truncated after the quadratic term.
It would be exact for the special case of a bar with cross-sectional area varying according to
\beq
A(x) = \frac{A_c}{1-\frac{x^2}{\lg^2}} =\frac{A_c\lg^2}{\lg^2-x^2} 
\eeq
where $A_c$ is the area of the weakest section and 
$\lg$ needs to be larger than the distance of that section from the bar end.
However, even for bars with a more general but smooth variation of cross-sectional area, the quadratic
stress distribution  (\ref{eqgp7}) is a good approximation of the actual one in the vicinity of the weakest
section, where the plastic zone is expected to develop. Therefore, analytical solutions of this special case
can be considered as representative of other cases with more general but still smooth stress variations.
At the same time, if the mathematical problem is interpreted as describing a bending beam instead of 
an axially loaded bar, the quadratic distribution of bending moments exactly corresponds to the fundamental
case of a simply supported beam of length $2\lg$ subjected to a uniform transversal load; see Fig.~\ref{fbeam}a.

The second case, a continuous variation of stress with discontinuous spatial derivative,
corresponds to  an axially stretched bar with a wide V-shaped notch;
see Fig.~\ref{fgp1}b. The  stress distribution to be used in the analytical solution is given by
\beql{eqgp9}
\sig(x) = \sigc \left(1-\frac{\vert x\vert}{\lg}\right)
\eeq
Again, for this relation to be exact, the notch would need to have a specific shape, with the sectional
area varying according to
\beql{eqgp9b}
A(x) = \frac{A_c}{1-\frac{\vert x\vert}{\lg}} = \frac{A_c\lg}{\lg-\vert x\vert}
\eeq
From a more general point of view,
the simple piecewise linear stress distribution serves as a prototype of all distributions
with a kink at the weakest section. In terms of the beam bending problem, it exactly corresponds
to another fundamental case of a simply supported beam of length $2\lg$ subjected to a concentrated
transversal force at midspan; see Fig.~\ref{fbeam}b.

For each type of stress distribution and each gradient plasticity model, the post-yield response will be
analyzed in terms of the 
distribution of plastic strain, evolution of plastic zone size, and load-displacement diagram.
This will provide insight into the interplay between the material length scale introduced by the 
gradient enhancement of the elastoplastic model and the geometric length scale related to the variation
of sectional area or, for the bending problem, to the span of the beam.

\section{Explicit second-order gradient plasticity model}

The explicit gradient plasticity model directly incorporates the second spatial gradient
of cumulative plastic strain into the softening law. In the case of linear softening and
in the one-dimensional setting,
equation (\ref{eqgp2a}) is replaced by
\begin{equation}
\sigY=\sig_0+H(\kap+l^2\kap'')
\label{eqgp11}
\end{equation}
where $l$ is a new material parameter with the dimension of length, and primes denote
derivatives with respect to the spatial coordinate $x$. In a general multidimensional setting,
the second spatial derivative would be replaced by the Laplace operator.

In the elastic range, the plastic strain
identically vanishes and the yield stress $\sigY$ is at its initial level, $\sig_0$. As long as the
stress $\sig(x)$ is everywhere below $\sig_0$, the response must remain elastic.
The onset of plastic yielding occurs when the stress in the weakest section, $\sigc$,
reaches the initial yield stress, $\sig_0$. For a standard (not enriched) softening model, 
plastic yielding would localize into this
single section. The gradient enhancement is supposed to act as a localization limiter,
which makes the plastic zone grow to a finite size $\Lp$.  This process will now be studied
analytically for the two basic types of stress distributions, (\ref{eqgp7}) and (\ref{eqgp9}). 
It can be expected that a contiguous plastic zone
forms around the weakest section and the surrounding parts of the bar remain elastic. 
Due to symmetry, the plastic zone is  assumed to be an interval $\Ip=(-\Lp/2,\Lp/2)$,
symmetric with respect to the origin.
Of course,
the boundaries between the elastic and plastic zones can move, but as long as the plastic zone
does not shrink (i.e., $\Lp$ grows monotonically), 
the elastic zone is characterized by zero plastic strain. In the plastic
zone, the yield function must vanish, which provides an equation for the determination of plastic strain.
For the gradient-enhanced model, this equation has a differential character.

\subsection{Quadratic stress distribution}
\label{sec:2.1}

At all points of the plastic zone, the current yield stress $\sigY$ given by (\ref{eqgp11})
must be equal to the stress $\sig$.
For the quadratic stress distribution (\ref{eqgp7}), this condition leads to the
equation
\begin{equation}
\kap(x)+l^2\kap''(x)=\frac{\sig_c-\sig_0-\sig_c x^2/\lg^2}{H}
\label{eqgp13}
\end{equation}
This is a nonhomogeneous second-order linear differential equation with constant coefficients, and its general
solution can be constructed as the sum of a particular solution of the  nonhomogeneous equation and the general solution
of the corresponding homogeneous equation. For the quadratic function on the right-hand side of (\ref{eqgp13}),
there exists a quadratic particular solution 
\beql{eqgp14}
\kap(x) = A_1 + A_2 x^2
\eeq
with constants $A_1$ and $A_2$ easily identified by substituting (\ref{eqgp14}) into the left-hand side 
of (\ref{eqgp13}) and comparing the constant terms and the quadratic terms on both sides. By adding the 
general solution of the homogeneous equation, which is a linear combination of $\cos(x/l)$ and $\sin(x/l)$,
the general solution of (\ref{eqgp13}) is obtained in the form 
\begin{equation}
\kap(x)=\frac{\sig_c(\lg^2+2l^2-x^2)-\sig_0\lg^2}{H\lg^2}+C_1\cos{\frac{x}{l}}+C_2\sin{\frac{x}{l}}
\label{eqgp14x}
\end{equation}
Integration constants $C_1$ and $C_2$ need to be determined from 
conditions $\kap=0$ and $\kap'=0$ imposed at the boundary of the plastic zone.
These are sometimes considered as boundary conditions, but they are better justified
by regularity requirements. 
The plastic strain identically vanishes outside the plastic zone,
and if it did not remain continuously differentiable across the elasto-plastic boundary, its
second derivative would have a singular character and the yield condition would be ``strongly violated''.
Precise mathematical justification of these statements, based on a variational formulation of the problem,
will be presented in a separate paper.
For the present purpose it is sufficient to admit that the plastic strain and its spatial derivative must vanish
at the boundary of the interval that corresponds to the plastic zone. The end points of this interval are not
known in advance, so in total we would have four unknowns (two integration constants and two coordinates
of the end points) and four conditions.

However, due to symmetry of the problem, the plastic zone $\Ip=(-\Lp/2,\Lp/2)$
is centered at the origin, and the plastic strain
distribution is described by an even function. This latter condition implies that integration constant $C_2$
must vanish. 
So it remains to determine $C_1$ and $\Lp$ from conditions
\begin{equation}
\kap(L_p/2)=0, \quad \kap'(L_p/2)=0
\label{eqgp15}
\end{equation}
Substituting the general solution (\ref{eqgp14x}) with $C_2=0$ into (\ref{eqgp15}) leads to 
two equations 
\bea
\sig_c(\lg^2+2l^2-\mbox{$\frac{1}{4}$}\Lp^2)-\sig_0\lg^2+C_1H\lg^2\cos{\frac{\Lp}{2l}}=0
\\
\sigc\Lp l+C_1H\lg^2\sin{\frac{\Lp}{2l}}=0
\label{eqgp16}
\eea
that are linear in terms of $C_1$ and nonlinear in terms of $\Lp$.  Unknown $C_1$ is easily 
eliminated and the resulting equation for unknown $\Lp$ reads 
\begin{equation}
\tan\frac{L_p}{2l}=\frac{\sigc L_pl}{\sigc(\lg^2+2l^2-\mbox{$\frac{1}{4}$}L_p^2)-\sig_0\lg^2}
\label{eq:tan1}
\end{equation}
Recall that parameter $\sig_0$ is a fixed material property while parameter $\sigc$ represents the current stress
in the weakest section and is directly related to the axial force transmitted by the bar, $F=A_c\sig_c$.
Plastic yielding starts when $\sigc=\sig_0$, i.e., when $F=F_0$ where $F_0=A_c\sig_0$ is the limit elastic force.
For a given axial force $F$, the corresponding size of the plastic zone $\Lp$ could be obtained 
by solving nonlinear equation (\ref{eq:tan1}) numerically, with $\sigc$ set to $F/A_c$. 
However, for some load levels there could be multiple solutions
or none at all. It is much more convenient to revert the procedure and express the axial force
in terms of the plastic zone size,
because this can be done analytically:
\beql{eqfac}
F=A_c\sigc = \frac{A_c\sig_0\lg^2}{\lg^2+2l^2-\frac{1}{4}\Lp^2-\Lp l\,\cotan(\Lp/2l)}
\eeq
A better understanding of the role of individual parameters can be gained if the problem is described
in terms of dimensionless quantities. The force or stress level is reflected by the load parameter
\beql{eq:phi}
\phi=\frac{F}{F_0}=\frac{\sigc}{\sig_0}
\eeq
which is equal to one at the onset of yielding. The length variables $\lg$ and $\Lp$ are taken relative
to the material length parameter $l$. Therefore we introduce dimensionless variables  
\begin{figure}%
\centering
\includegraphics[scale=0.7]{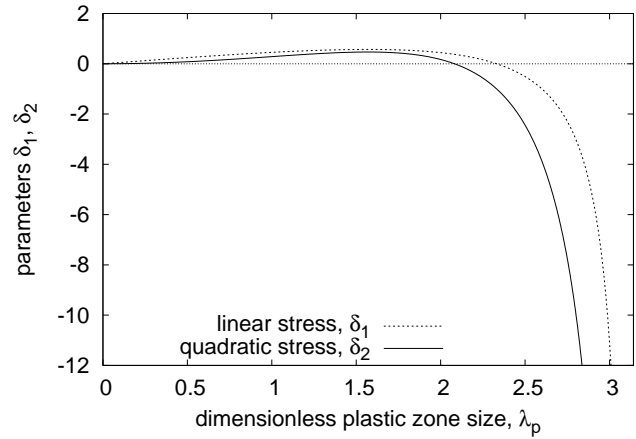}%
\caption{Explicit second-order
gradient plasticity model: 
Dependence between auxiliary parameters $\delta_2$ and $\delta_1$, respectively given by (\ref{eqgpdelta2})
and (\ref{eq:delta1}), and the dimensionless
plastic zone size $\lamp=\Lp/2l$}%
\label{fig:A2_root}%
\end{figure}
\begin{figure}%
\centering
\includegraphics[scale=0.7]{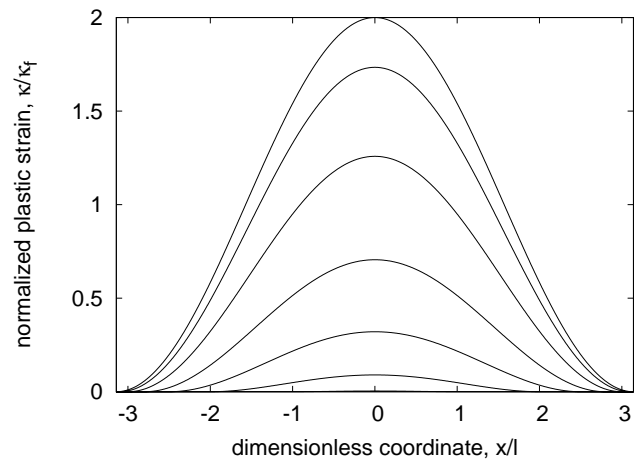}%
\caption{Explicit second-order
gradient plasticity model, quadratic stress distribution: Evolution of plastic strain profile}%
\label{fig:A2q_progress_of_kappa}%
\end{figure}
\beq
\lamp=\frac{L_p}{2l}, \hskip 5mm 
\lamg=\frac{\lg}{l}
\label{eqgp20}
\eeq
Variable $\lamp$ is the ratio between one half of the plastic zone size and the characteristic material length.
Factor 2 in the denominator simplifies the results, because the boundary of the plastic zone is
at $x=l\lamp$. Parameter $\lamg$ is the ratio between the geometric and material characteristic
length scales. In terms of the dimensionless variables, equation (\ref{eqfac}) reads
\beql{eq:phi1}
\phi= \frac{\lamg^2}{\lamg^2+2-\lamp^2-2\lamp \,\cotan\lamp}= \frac{\lamg^2}{\lamg^2-\delta_2}
\eeq
where
\beql{eqgpdelta2}
\delta_2 = \lamp^2+2\lamp\,\cotan\lamp-2
\eeq
is an auxiliary variable introduced for convenience---it permits isolating the effect of $\lamg$ from the dependence
between $\lamp$ and $\phi$. 
The relation between $\lamp$ and $\delta_2$ is graphically represented by the solid curve in Fig.~\ref{fig:A2_root}.
Parameter $\delta_2$ continuously grows from its initial value $0$ at $\lamp=0$ to its maximum value
$\delta_{2,\max{}}=\pi^2/4-2$ attained at $\lamp=\pi/2$, and afterwards decreases and tends to minus infinity as $\lamp$
approaches $\pi$ from the left. The corresponding load parameter $\phi$ (axial force transmitted by the bar
normalized by its elastic limit value) continuously grows
from $\phi_0=1$ at the elastic limit to its maximum value
\begin{equation}
\phi_{\max{}} = \frac{F_{\max{}}}{F_0}=\frac{\lamg^2}{\lamg^2+2-\pi^2/4}
\label{eq26}
\end{equation}
and afterwards decreases to zero as $\lamp$ approaches $\pi$ from the left. An important point is that, during
this process, $\lamp$ grows monotonically, and so the assumption of plastic zone expansion is verified and
the solution is admissible. The loading process can be parameterized by $\lamp$ ranging from $0$ to $\pi$.

\begin{figure}%
\centering
\begin{tabular}{cc}
(a) & (b)
\\
\includegraphics[scale=0.7]{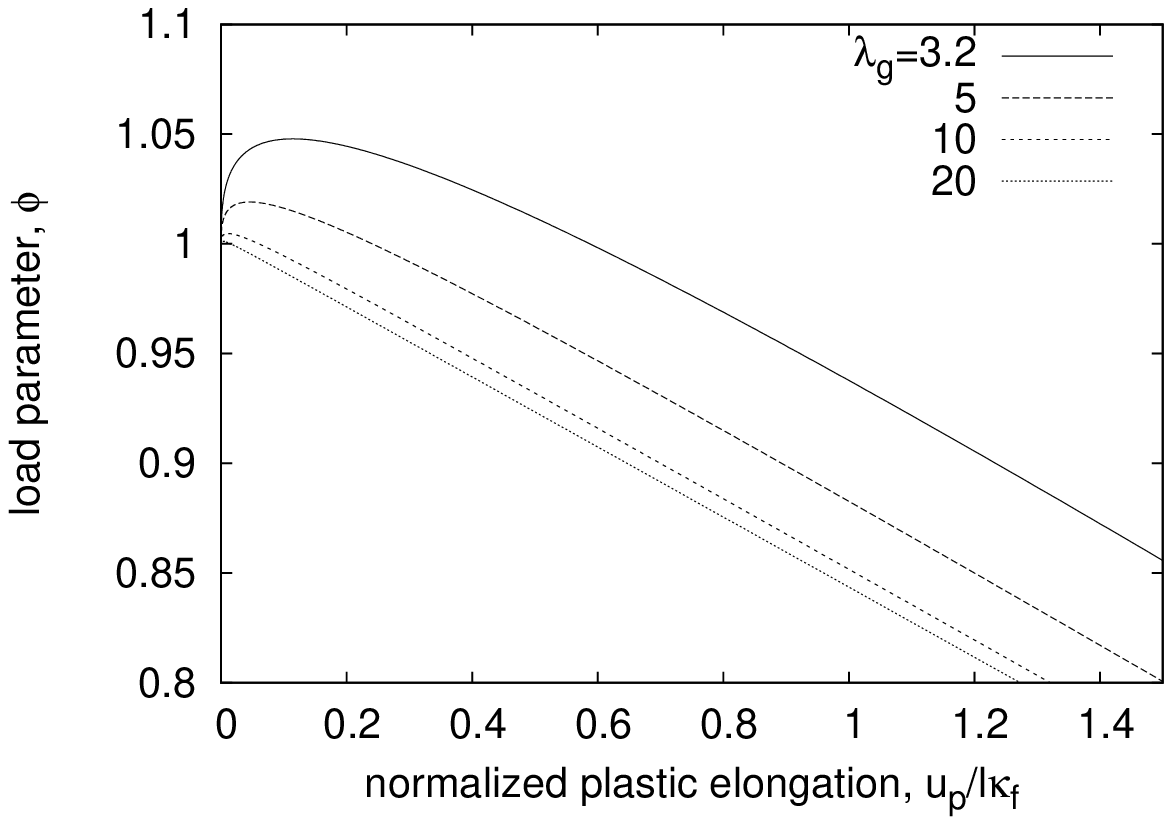}%
&
\includegraphics[scale=0.7]{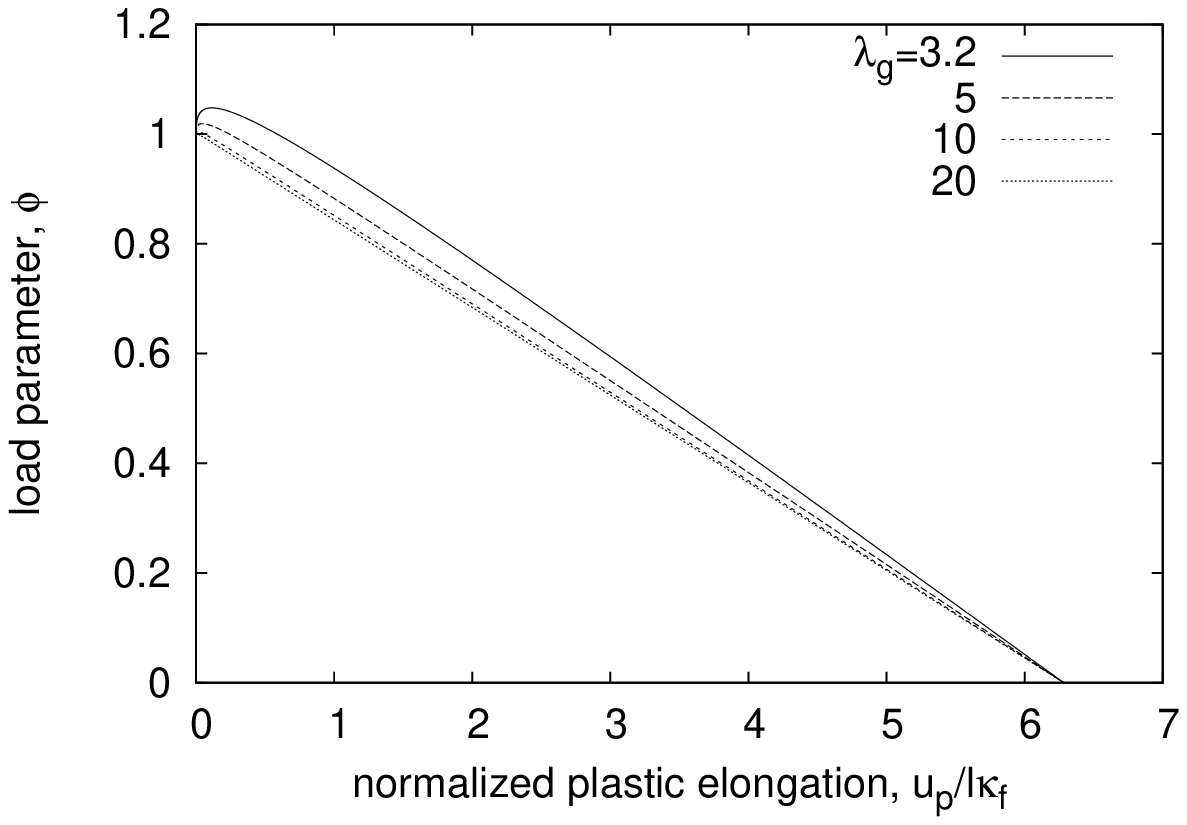}%
\end{tabular}
\caption{Explicit second-order
gradient plasticity model, quadratic stress distribution: Plastic part of load-displacement diagram --- (a) close-up of the initial part, (b) complete diagram }%
\label{fig:A2q_stress-strain_diagram}%
\end{figure}

For each $\lamp\in[0,\pi)$ and the corresponding $\sigc=\phi\sig_0$ determined from (\ref{eq:phi1}),
integration constant $C_1$ can be computed from (\ref{eqgp16}) and the resulting expression
\beq
C_1=-\frac{2\sig_cl^2\lamp}{H\lg^2\sin\lamp}
\eeq
can be substituted into the general solution (\ref{eqgp14x}). This gives the function
\begin{equation}
\kap(x)=\frac{1}{H\lg^2}\left[\sig_c(\lg^2+2l^2-x^2)-\sig_0\lg^2-
\frac{2\sig_cl^2\lamp}{\sin\lamp}\cos{\frac{x}{l}}\right]
\label{eq:A2q_kappa}
\end{equation}
describing the plastic strain distribution along the process zone. 
For the representation of the plastic strain profile, it is again convenient to use normalized variables.
The plastic strain is already dimensionless, but it is useful to normalize it by the reference value $\kappa_f=-\sig_0/H$,
which corresponds to the plastic strain at full softening to zero stress according to the local linear softening law.
The spatial coordinate is naturally normalized by the material characteristic length $l$, which leads to the
dimensionless coordinate $\xi=x/l$. The distribution of normalized plastic strain, $\kapt=\kap/\kap_f$,
is then described by the function
\beql{eq:31}
\kappat(\xi)=\frac{\kap(l\xi)}{\kappa_f}=1-\phi+\frac{\phi}{\lamg^2}\left(\frac{2 \lamp}{\sin\lamp}\cos{\xi}+\xi^2-2\right)
\eeq
where $\phi$ depends on $\lamp$ and $\lamg$ according to (\ref{eq:phi1}). 
For a fixed ratio $\lamg=\lg/l$ and a sequence of
dimensionless plastic zone sizes $\lamp$, equation (\ref{eq:31}) provides a sequence of plastic strain profiles. 
An example with $\lamg=5$ and $\lamp=1$, 2, 2.5, 2.8, 3, 3.1 and $\pi$
is plotted in Fig.~\ref{fig:A2q_progress_of_kappa}. It is clear that at each point 
the plastic strain evolves monotonically, which verifies
the admissibility of our solution. 

Finally, integrating the plastic strain along the entire process zone,
we obtain 
\beql{eqgpup}
u_p = \int_{-\Lp/2}^{\Lp/2}\kappa(x)\,\dx = \frac{\Lp}{H}\left(\sigc-\sig_0-\frac{\sigc\Lp^2}{12\lg^2}\right)
\eeq
Variable $u_p$ can be interpreted as the plastic part of the bar elongation. The total bar elongation is
the sum of $u_p$ and the elastic elongation $u_e=C_eF$ where
\beql{eq31}
C_e=\int_{\cal L}\frac{\dx}{EA(x)}
\eeq
is the elastic compliance of the bar. In (\ref{eq31}), 
${\cal L}$ denotes the one-dimensional domain representing the entire bar
(interval of length $L$).  The elastic compliance depends on the total bar length $L$, and thus the overall load-displacement
diagram in terms of $F$ against $u_e+u_p$ would also be affected by the bar length. Since the plastic part
of elongation, $u_p$, is independent of $L$ (as long as the size of the plastic zone does not exceed the bar length,
which could happen only for extremely short bars), the plastic part of the load-displacement diagram,
expressed in terms of $F$ against $u_p$, depends only on parameters $l$, $\lg$, $H$, $\sig_0$ and $A_c$. 
If we convert it to the dimensionless
representation, with $u_p$ divided by the positive constant $l\kappa_f$ and $F$ divided by the elastic limit force $F_0=\sig_0A_c$,
we obtain
\beql{eqgpup1}
\frac{u_p}{l\kappa_f} = 2\lamp\left(1-\phi+\frac{\phi\lamp^2}{3\lamg^2}\right) 
\eeq
For a fixed value of the dimensionless parameter $\lamg=\lg/l$,  equations (\ref{eq:phi1}) and (\ref{eqgpup1}) provide a parametric
description of the plastic part of the load-displacement diagram, parameterized by $\lamp$.
The geometric length $\lg$ is typically much larger than the material length $l$.
In fact, for the above solution to be valid in the entire softening range, $\lg$ should be larger than one half of the
maximum plastic zone size, $\pi l$, otherwise the boundary of the plastic zone would reach the end of the bar and
the regularity conditions at $x=\pm\Lp/2$ should be replaced by boundary conditions at $x=\pm L/2$. Therefore, 
the diagrams in Fig.~\ref{fig:A2q_stress-strain_diagram} are plotted for $\lamg=20$, 10, 5 and 3.2.
It is found that, in accordance with what was already
seen in Fig.~\ref{fig:A2_root}, in the first stage of the yielding process the force transmitted by the bar continuously increases from
the elastic limit value $F_0$ to the maximum value $F_{\max{}}=\phi_{\max{}}F_0$ and only later starts decreasing and eventually vanishes. Note that the limit $\lamg\rightarrow\infty$ corresponds to a bar with a uniform section, for which the
plastic zone size jumps immediately to $2\pi l$ and the global response does not exhibit hardening at all.

Summarizing the results obtained in this subsection for the smooth nonuniform
distribution of stress and second-order explicit gradient plasticity model with linear softening, it can be concluded that
plastic yielding initiates in the weakest cross section (with the maximum stress) and then progressively expands,
first at {\bf increasing axial force}. Even though the local constitutive law assumes softening right from the
onset of plastic yielding, the contribution of the gradient enhancement makes the yield stress in the weakest
section and its neighborhood grow, until a certain size of the process zone is reached. The maximum force is therefore
higher than the limit elastic force, and the global load-displacement diagram exhibits hardening. 
From (\ref{eq26}) it follows that
\beq
\frac{1}{F_0}-\frac{1}{F_{\max{}}}=\frac{\pi^2/4-2}{F_0\lamg^2}=\frac{\pi^2/4-2}{F_0}\left(\frac{l}{\lg}\right)^2
\eeq
which means that the difference between the reciprocal value
of the limit elastic load (i.e.\ the ultimate
load for the standard model) and the 
reciprocal value
of the actual ultimate load for the gradient model   is proportional to the square of the ratio between
the material length parameter $l$ and       the geometric length parameter $\lg$. This observation provides some
insight into the interplay between the length scales of the problem. For $l$ approaching zero, the hardening effects
fade away because the gradient regularization is suppressed and the model approaches a standard one.
For $\lg$ approaching infinity, the hardening effect fades away because the bar approaches a prismatic one,
with uniform stress distribution, for which the global response exhibits linear softening without any previous
hardening and the uniform response bifurcates into solutions with plastic strain localized into an interval
of size $2\pi l$, the position of which is completely arbitrary. For the nonuniform bar with quadratic stress
distribution, $2\pi l$ is the limit that the process zone size approaches as the axial force transmitted by the bar
decreases to zero during the second stage of failure, characterized by global softening. 

The same discussion applies to problems of beam bending with quadratic distribution of bending moment,
e.g., to the simply supported beam loaded by uniform transversal load; see Fig.~\ref{fbeam}a. 
The middle section transmitting the largest moment
 starts failing first, and for a standard model with softening incorporated into the moment-curvature relation,
inelastic processes would localize into this single section and the beam would fail at zero dissipation.
For the second-order explicit gradient model, with the current yield moment given by
\beq
M_Y = M_0 + C(\kappa_p+l^2\kappa_p'')
\eeq
plastic yielding spreads to a finite segment of the beam, corresponding to an inelastic hinge. Note that
the plastic part of curvature, $\kappa_p$, is now marked with a subscript $p$, because $\kappa$ is used
for the total curvature, consisting of the elastic and plastic parts. The integral of $\kappa_p$ along the plastic
zone corresponds to the rotation of an equivalent inelastic hinge concentrated into a single cross section,
same as $u_p$ in equation (\ref{eqgpup}) can be interpreted as the opening of an equivalent cohesive crack. The inelastic 
hinge is therefore first hardening and only later softening. The peak bending moment depends on the geometric
length scale, in the case of beam bending on the beam span, which introduces a size effect: shorter beams appear
to be stronger (in terms of the maximum bending moment that they can transmit). 

Solution of one half of the simply 
supported beam also applies to a cantilever, provided that a homogeneous Neumann boundary condition for the
plastic curvature is enforced at the clamped support. The cantilever problem was analyzed in \cite{Cha08},
with the conclusion that the explicit gradient model is inappropriate and cannot describe the evolution of the
plastic hinge. However, this conclusion was based on the tacit assumption that the global response
must be softening right from the onset of plastic yielding. The present analysis shows that a reasonable solution
can be found, with a continuous evolution
of all quantities and a very reasonable physical interpretation, once it is accepted that, under nonuniform stress,
local softening does not necessarily imply global softening. It is quite natural that, for a gradient model
that exhibits (at least in a weak sense) a nonlocal behavior, the strength of the structure does not depend
exclusively on the weakest section but is influenced by the distribution of local strength in a neighborhood of that
section, the size of which is controlled by the material characteristic length. Here we mean strength in the broad
sense, in the present example referring to the axial force or bending moment that can be transmitted by a cross section. 
We have considered a variation of this sectional strength due to a nonuniform sectional area, but qualitatively
the same results would be obtained in the case of a nonuniform local material strength (initial
yield stress $\sig_0$).

\subsection{Piecewise linear stress distribution}

Now we proceed to the piecewise linear stress distribution from Fig.~\ref{fgp1}b,
with a discontinuous spatial derivative at the weakest section. The analysis proceeds along similar lines
as for the quadratic stress distribution in the previous subsection, and so we omit most of the detailed  explanatory
comments. Whenever possible, the solution will be presented in terms of dimensionless quantities.

In the plastic zone, the stress given by (\ref{eqgp9}) must be equal to the yield stress given by (\ref{eqgp11}),
which leads to the differential equation
\begin{equation}
\kap(x)+l^2\kap''(x)=\frac{\sig_c-\sig_0}{H}-\frac{\sig_c\vert x\vert}{H\lg}
\end{equation}
or, in dimensionless form,
\begin{equation}
\kapt(\xi)+\kapt''(\xi)=1-\phi+\phi\lamg^{-1}\vert\xi\vert
\label{eq:32}
\end{equation}
where $\kapt=\kap/\kap_f$, $\kap_f=-\sig_0/H$, $\phi=\sigc/\sig_0$, $\lamg=\lg/l$ and,
for simplicity, the derivative with respect to the dimensionless coordinate $\xi=x/l$ is denoted by a prime,
even though before the prime was used for the derivative with respect to $x$.
The general solution of (\ref{eq:32}) is
\begin{equation}
\kapt(\xi)=\left\{\begin{array}{ll}
1-\phi-\phi\lamg^{-1}\xi+C_1\cos{\xi}+C_2\sin{\xi} & \mbox{ for } -\lamp\le \xi\le 0
\\[3mm]
1-\phi+\phi\lamg^{-1}\xi+C_3\cos{\xi}+C_4\sin{\xi} & \mbox{ for } 0\le \xi\le \lamp
\end{array}\right.
\label{eqgp31}
\end{equation}
Continuity at $\xi=0$ implies that $C_1=C_3$, and
continuous differentiability at $\xi=0$ combined with symmetry  implies that 
\beq
C_2=-C_4=\phi\lamg^{-1}
\eeq
So the general solution (\ref{eqgp31}) can be rewritten as
\beql{eqgp31z}
\kapt(\xi)=1-\phi+\phi\lamg^{-1}\left(\vert\xi\vert-\sin\vert\xi\vert\right) +C_1\cos\xi
\eeq
Integration constant $C_1$ and the size of the plastic zone $\Lp$ are determined from the conditions 
$\kapt(\lamp)=0$ and $\kapt'(\lamp)=0$. Elimination of the integration constant leads to a nonlinear equation
\beql{eqgp31x}
\tan\lamp = \frac{\phi\left(1-\cos\lamp\right)}{(\phi-1)\lamg+\phi\left(\sin\lamp-\lamp\right)}
\eeq
that links the plastic zone size $\lamp$ to the load parameter $\phi$, with an influence of parameter $\lamg$.

The dependence on $\lamg$ can be treated separately if we define an auxiliary parameter 
\beql{eqgp32}
\delta_1 = \left(1-\frac{1}{\phi}\right)\lamg
\eeq
This parameter can be expressed from (\ref{eqgp31x}) 
as a function of the dimensionless plastic zone size:
\beq
\delta_1=\lamp-\sin\lamp+\frac{1-\cos\lamp}{\tan\lamp}=\lamp-\tan\frac{\lamp}{2}
\label{eq:delta1}
\end{equation}
The corresponding graph is represented by the dashed curve in Fig.~\ref{fig:A2_root}.
As $\lamp$ grows,
parameter $\delta_1$ first increases from
its initial value $\delta_1=0$ at $\lamp=0$ to its maximum $\delta_{1,\max{}}=\pi/2-1$
at $\lamp=\pi/2$, and then decreases and tends to minus infinity as
$\lamp$ approaches $\pi$ from the left. This means that, similar to the previous case
of quadratic stress distribution, the axial force  transmitted
by the bar,
\beql{eqxx}
F=F_0\phi=\frac{F_0}{1-\delta_1\lamg^{-1}}=\frac{F_0\lamg}{\lamg-\lamp+\tan(\lamp/2)}
\eeq
 is first increasing, even after the onset of plastic yielding,
and the global response is hardening up to the peak force 
\begin{equation}
F_{\max{}}=\frac{F_0\lamg}{\lamg-\delta_{1,\max{}}}=\frac{F_0\lamg}{\lamg+1-\pi/2}
\end{equation}
After that the force decreases to zero and the plastic zone keeps expanding
up to its maximum size $2\pi l$.

\begin{figure}%
\centering
\begin{tabular}{cc}
(a) & (b)
\\
\includegraphics[scale=0.7]{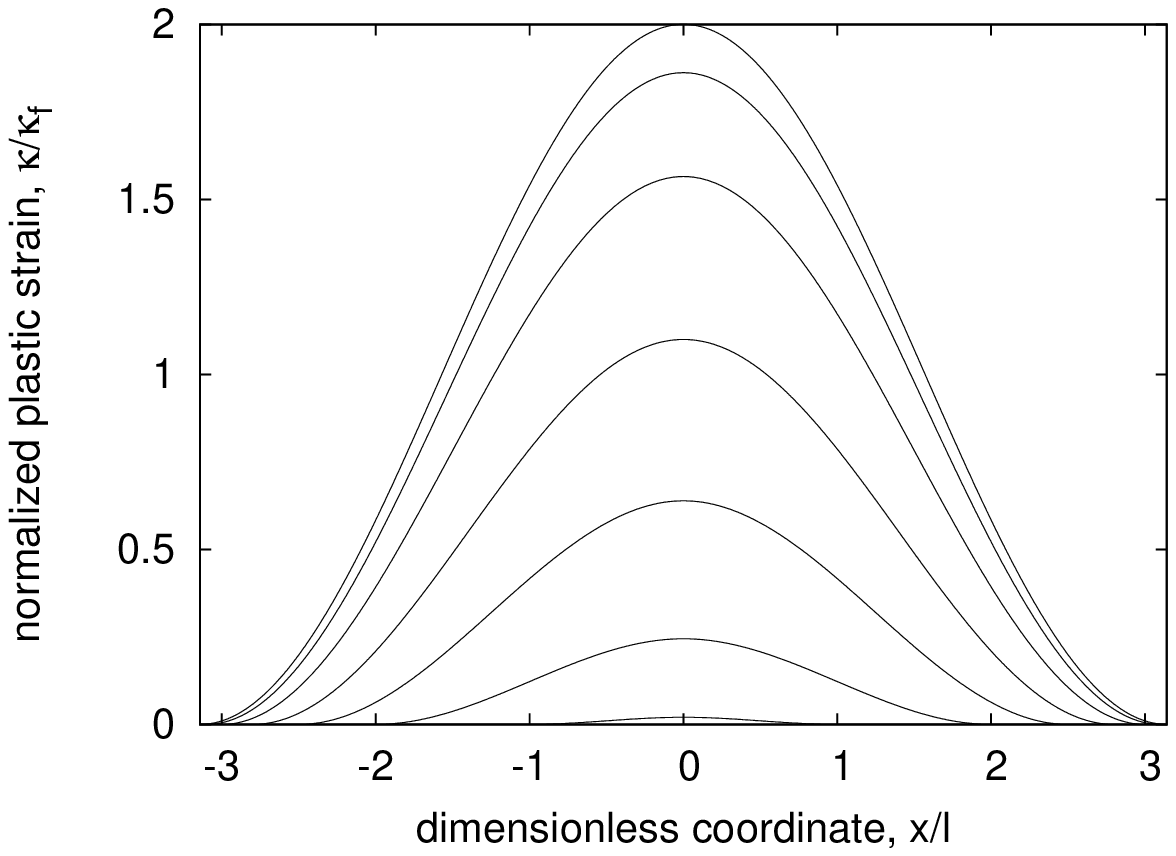}%
&
\includegraphics[scale=0.7]{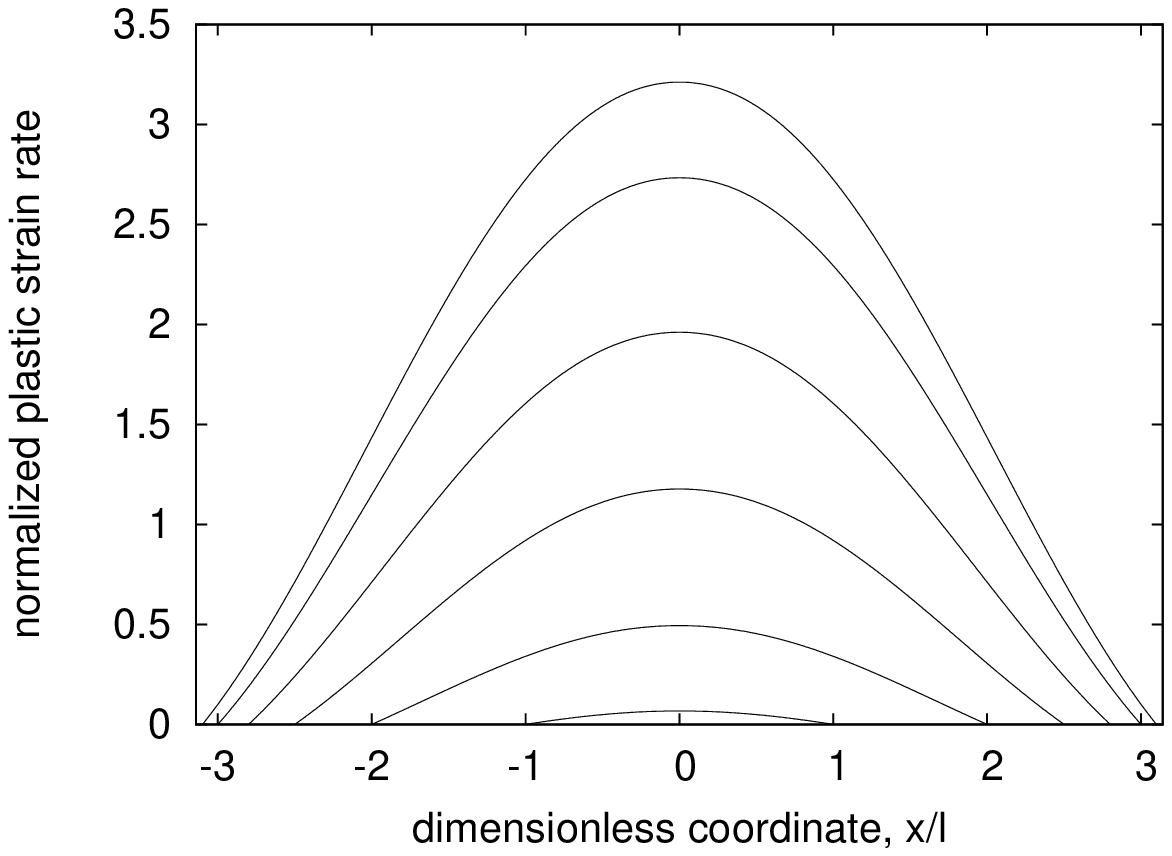}%
\end{tabular}
\caption{Explicit second-order
gradient plasticity model, piecewise linear stress distribution: Evolution of (a) plastic strain and (b) plastic strain rate}%
\label{fig:A2l_progress_of_kappa}%
\end{figure}

\begin{figure}%
\centering
\begin{tabular}{cc}
(a) & (b)
\\
\includegraphics[scale=0.7]{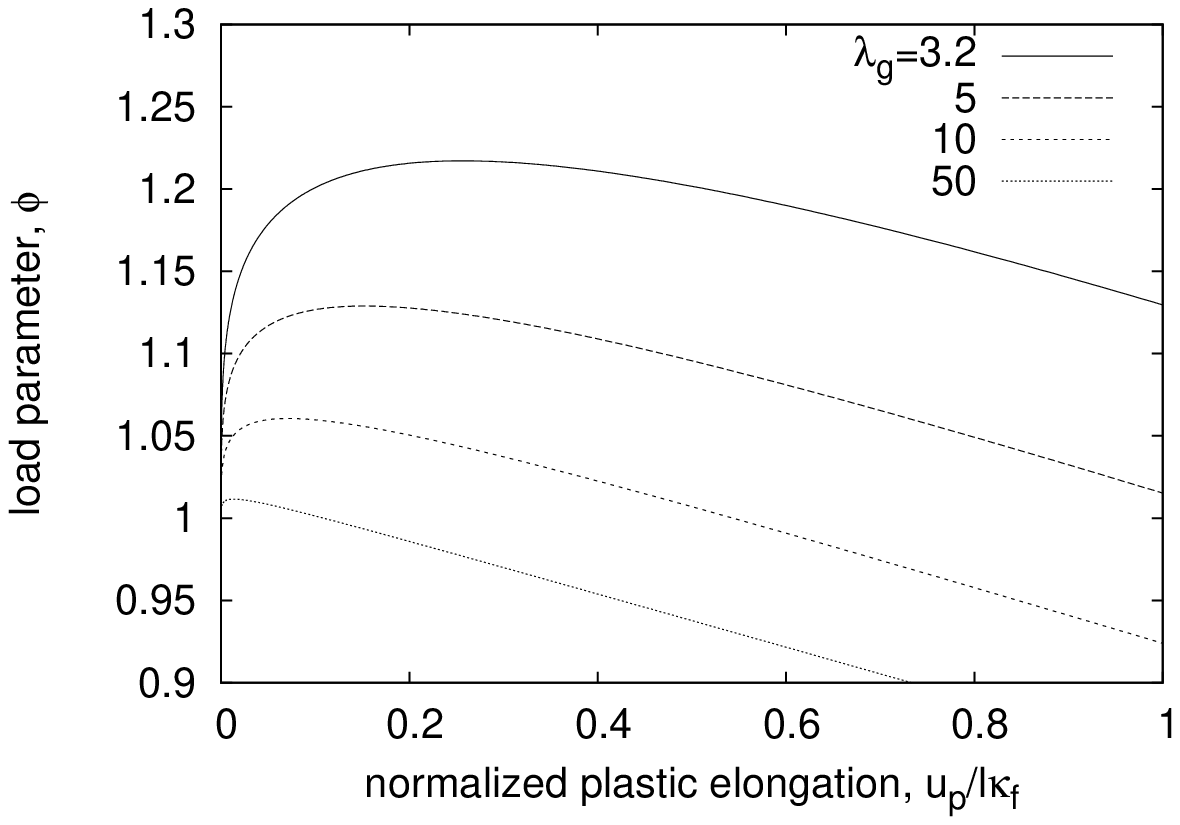}%
&
\includegraphics[scale=0.7]{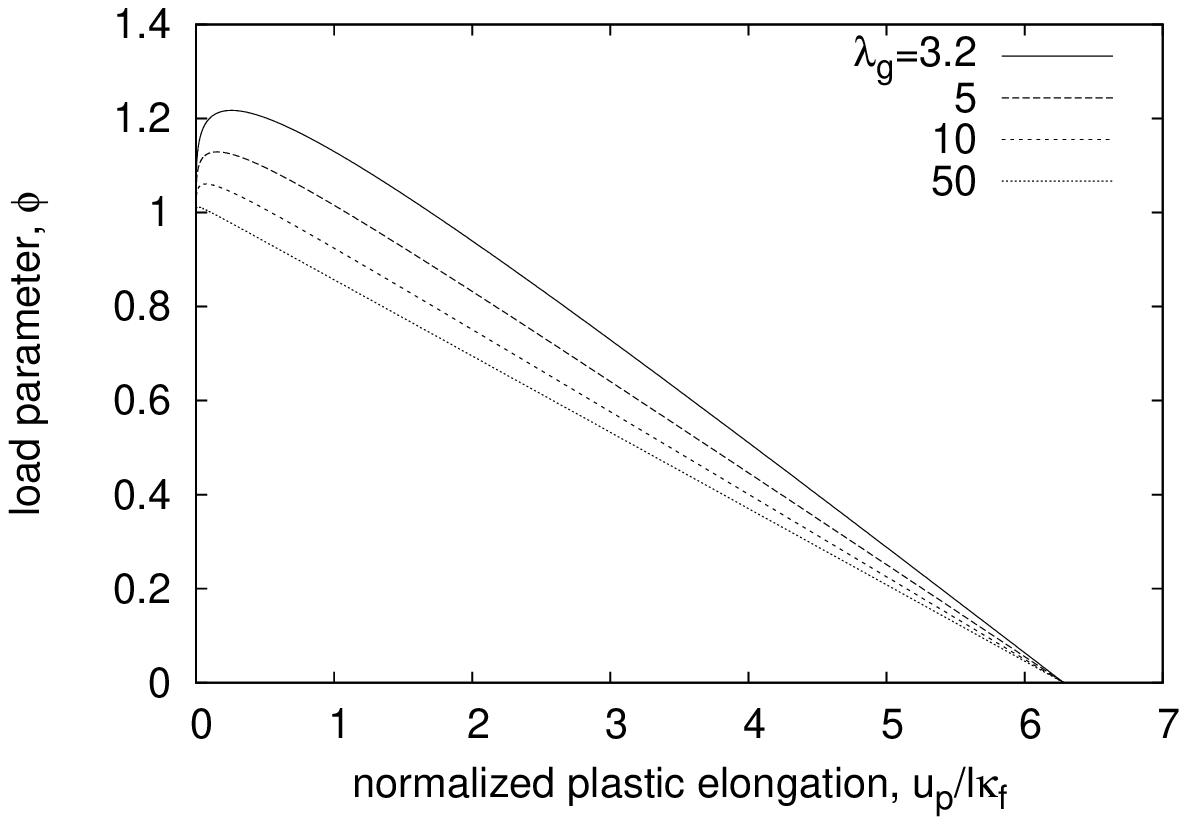}%
\end{tabular}
\caption{Explicit second-order
gradient plasticity model, piecewise linear stress distribution: Plastic part of load-displacement diagram --- (a) close-up of the initial part, (b) complete diagram }%
\label{fig:A2l_stress-strain_diagram}%
\end{figure}

Integration constant $C_1$ can be expressed as
\beq
C_1=
\phi\lamg^{-1}\tan\frac{\lamp}{2}
\eeq
and substituted into (\ref{eqgp31z}), which leads to the particular solution 
\begin{equation}
\kapt(\xi)=1-\phi
+\phi\lamg^{-1}\left(\vert \xi\vert-\sin\vert \xi\vert
+\tan\frac{\lamp}{2}\cos\xi\right)
\label{eq:A2l_kappa}
\end{equation}
The normalized plastic displacement is then
\begin{equation}
\frac{u_p}{l\kap_f}=\int_{-\lamp}^{\lamp} \kapt(\xi)\,\dxi = 2\lamp\left(1-\phi+\frac{\phi\lamp}{2\lamg}\right)
\end{equation}
The evolution of the plastic strain profile is shown in Fig.~\ref{fig:A2l_progress_of_kappa}a
for $\lamg=5$. Same as for the quadratic stress distribution,
the plastic strain grows monotonically at each fixed spatial point, which verifies the admissibility of the solution.
The plastic part of the load-displacement diagram is plotted in Fig.~\ref{fig:A2l_stress-strain_diagram}
for several values of $\lamg$.  

It is also instructive to look at the distribution of the plastic strain rate.
In equation (\ref{eq:A2l_kappa}), $\lamg$ is a fixed parameter and $\xi$ is the spatial coordinate.
Only the plastic zone size $\lamp$ and the load parameter $\phi$ evolve in time. 
Their rates are linked by the rate form of equations (\ref{eqgp32})--(\ref{eq:delta1}), from which
\beql{eq46a}
\dot{\phi}=\frac{\phi^2\cos\lamp}{2\lamg\cos^2\frac{\lamp}{2}}\,\dot{\lambda}_p
\eeq
Differentiating  (\ref{eq:A2l_kappa}) with respect to time and using (\ref{eq46a}), we obtain
\beq
\dot{\kappa}_n(\xi)= \left[\left(\vert \xi\vert-\sin\vert \xi\vert
+\tan\frac{\lamp}{2}\cos\xi-\lamg\right)\phi\cos\lamp+\lamg\cos\xi\right]
\frac{\phi\dot{\lambda}_p}{2\lamg^2\cos^2\frac{\lamp}{2}}
\eeq
It is possible to show that $\dot{\kappa}_n(\lamp)=0$ but  $\dot{\kappa}_n'(\lamp)\ne 0$.
This means that the plastic strain rate is continuous but not continuously differentiable (in space)
at the boundary of the plastic zone; see Fig.~\ref{fig:A2l_progress_of_kappa}b.
Therefore, the conditions of vanishing spatial derivative of plastic strain and of  
vanishing spatial derivative of plastic strain rate are in general not equivalent.
Only in the special case of a plastic zone of constant width, which occurs in a bar
with perfectly uniform properties, one can freely choose between conditions
 $\kappa_n'(\lamp)=0$ and   $\dot\kappa_n'(\lamp)=0$. 
The role of moving boundaries  of the plastic zone 
has been examined for instance in \cite{Pee07}.

\section{Explicit fourth-order gradient plasticity model}

Modified versions of the second-order 
model from the previous section can incorporate higher-order gradients
\cite{Zbib88,Muhlhaus91}.
A prototype model of this kind, using a fourth-order enrichment, postulates the softening law
in the form
\begin{equation}
\sig_Y=\sig_0+H\left(\kap-l^4\kap^{IV}\right)
\label{eqgp61}
\end{equation}
where superscript ``$IV$'' denotes the fourth spatial derivative. In a general multiaxial setting this would
be the ``Laplacean of the Laplacean''.

\subsection{Quadratic stress distribution}

For the quadratic stress distribution (\ref{eqgp7}), 
the distribution of plastic strain inside the plastic zone $\Ip=(-\Lp/2,\Lp/2)$
is governed by the fourth-order differential equation
\begin{equation}
\kap(x)-l^4\kap^{IV}(x)=\frac{\sig_c-\sig_0-\sig_cx^2/\lg^2}{H}
\label{eqgp62o}
\end{equation}
which can be obtained by combining 
the yield condition $\sig=\sigY$ with the softening law (\ref{eqgp61}).
In the dimensionless format, we rewrite it as
\begin{equation}
\kapt(\xi)-\kapt^{IV}(\xi)=1-\phi+\frac{\phi\xi^2}{\lamg^2}
\label{eqgp62}
\end{equation}
and construct the general solution
\begin{equation}
\kapt(\xi)=1-\phi+\frac{\phi\xi^2}{\lamg^2}+C_1\cos\xi+C_2\sin\xi+C_3\cosh\xi+C_4\sinh\xi
\label{eqgp63}
\end{equation}
where, as usual, $\kapt=\kap/\kap_f=-\kap H/\sig_0$, $\xi=x/l$, $\phi=\sigc/\sig_0$ and $\lamg=\lg/l$.

Integration constants $C_2$ and $C_4$ must vanish because of symmetry. 
The remaining unknowns are integration constants $C_1$ and $C_3$ and the size of the plastic zone $\Lp=2l\lamp$,
and they need to be determined from regularity conditions at the boundary of the plastic zone, i.e., at point
$\xi=\lamp$. 
Since we have only three unknowns, we cannot impose continuous differentiability up to the third order,
because this would represent four independent conditions. One may think that the discrepancy is caused by the
assumption of symmetry. This is not the case---if symmetry is not imposed, we have to determine four integration
constants and two coordinates of the boundary of the plastic zone (left and right boundary), which makes
a total of six unknowns, and again only three independent conditions can be satisfied at each boundary point. 
Consequently, continuity at the boundary of the plastic zone can be enforced only for $\kapt$ itself and its first and second derivative. 
As long as the plastic zone expands monotonically, the plastic strain is identically equal to zero outside 
the plastic zone, and
the continuity conditions read
\begin{equation}
\kapt(\lamp)=0, \quad \kapt'(\lamp)=0, \quad \kapt''(\lamp)=0
\label{eqgp64}
\end{equation}
The third derivative can be discontinuous, which means that the third derivative from the right (equal to zero)
and the third derivative from the left, denoted as $\kapt'''(\lamp^-)$, can be different. However, the jump
in the third derivative, given by $[[\kapt''']]_{\xi=\lamp}=0-\kapt'''(\lamp^-)=-\kapt'''(\lamp^-)$, 
must be nonnegative,\footnote{If the jump of the third derivative of plastic strain is positive, the 
fourth derivative contains a singular component that has
the character of a positive multiple of the Dirac distribution. Since the plastic modulus $H$ is negative,
 the yield stress at the boundary of the plastic zone computed from (\ref{eqgp61})
is ``positively infinite'' and the solution remains
plastically admissible. A precise justification of these rather intuitive arguments can be based
on the reformulation of the problem as a variational inequality.}
and so $\kapt'''(\lamp^-)$ must be nonpositive. This condition is to be verified once the solution is known.

After substitution of the general solution (\ref{eqgp63}), conditions  (\ref{eqgp64}) lead to the set of equations
\bea
\label{eq:cond4q_1}
C_1\cos{\lamp}+C_3\cosh\lamp  &=& -1+\phi-\phi\lamp^2\lamg^{-2}\\
\label{eq:cond4q_2}
-C_1\sin{\lamp}+C_3\sinh\lamp &=& -2\phi\lamp\lamg^{-2}\\
\label{eq:cond4q_3}
-C_1\cos{\lamp}+C_3\cosh\lamp &=& -2\phi\lamg^{-2}
\eea
Elimination of integration constants $C_1$ and $C_3$ reduces the problem to a single nonlinear equation for $\lamp$.
Using an auxiliary parameter 
\beq
\delta_2=\left(1-\frac{1}{\phi}\right)\lamg^2
\eeq
the resulting equation can be presented in the
dimensionless form
\begin{equation}
\tan\lamp=\frac{4\lamp-\left(2+\lamp^2-\delta_2\right)\tanh\lamp}{2-\lamp^2+\delta_2}
\end{equation}
In a similar spirit as in Section~\ref{sec:2.1}, 
instead of solving for $\lamp$, we can invert the problem and express 
\begin{equation}
\delta_2=\frac{4\lamp+(\lamp^2-2)\tan\lamp-(\lamp^2+2)\tanh\lamp}{\tan\lamp-\tanh\lamp}
\label{eq:4qtanfx}
\end{equation}
from which 
the load parameter is easily calculated as
\beql{eq58}
\phi=\frac{\lamg^2}{\lamg^2-\delta_2}
\eeq
The advantage is that, using this representation, the influence of parameters $\lamp$ and $\lamg$ 
is treated separately.

\begin{figure}%
\centering
\includegraphics[scale=0.7]{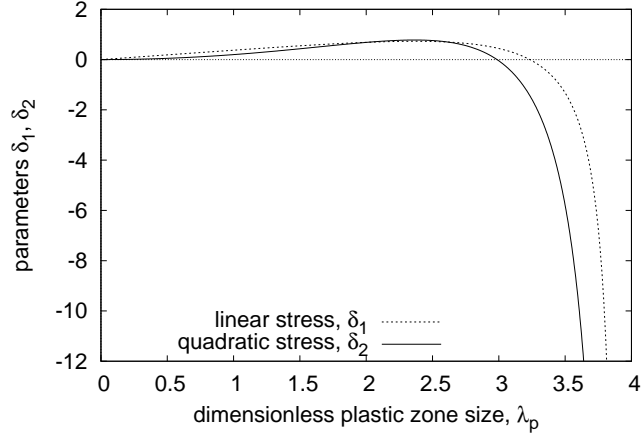}%
\caption{Explicit fourth-order gradient model: Dependence between dimensionless load parameters $\delta_1$ and $\delta_2$ and dimensionless plastic zone size $\lamp$}
\label{fig:comp_dl}%
\end{figure}

\begin{figure}
\centering
\begin{tabular}{cc}
(a) & (b) 
\\
\includegraphics[scale=0.7]{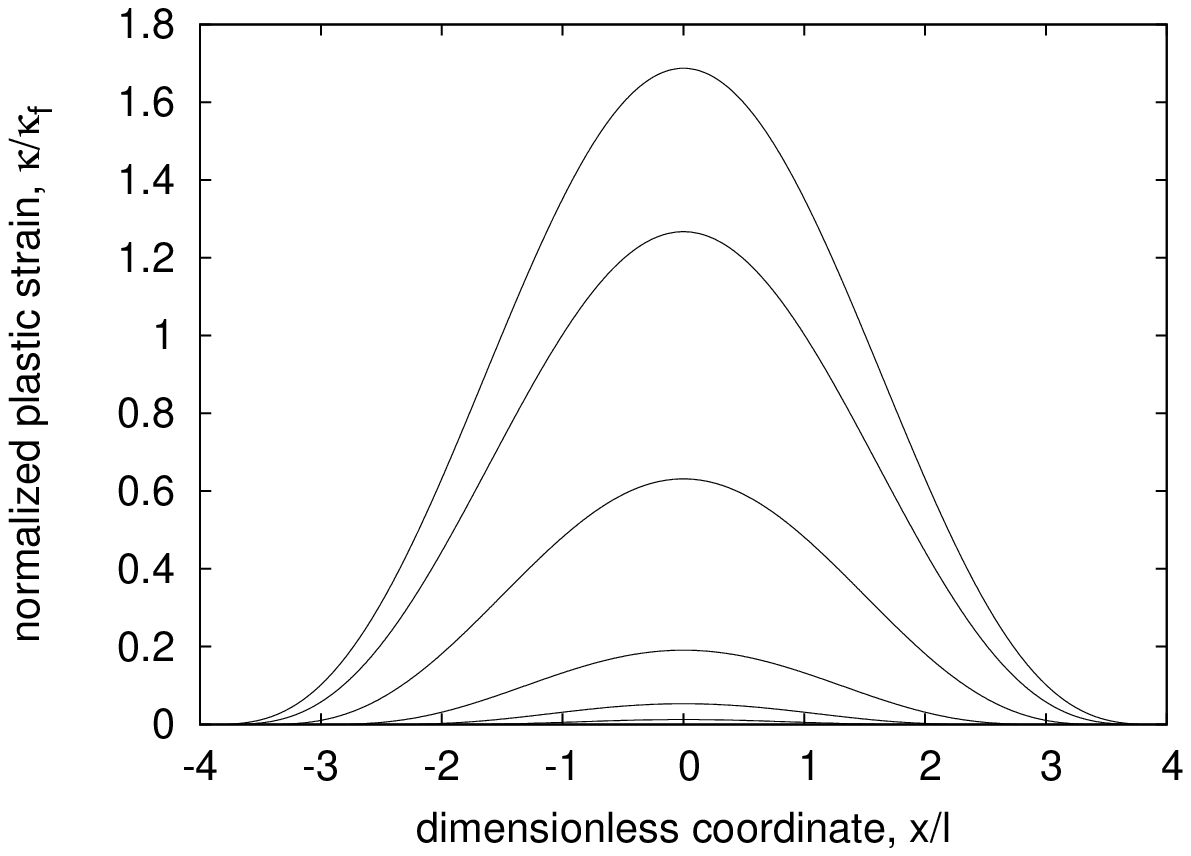}%
&
\includegraphics[scale=0.7]{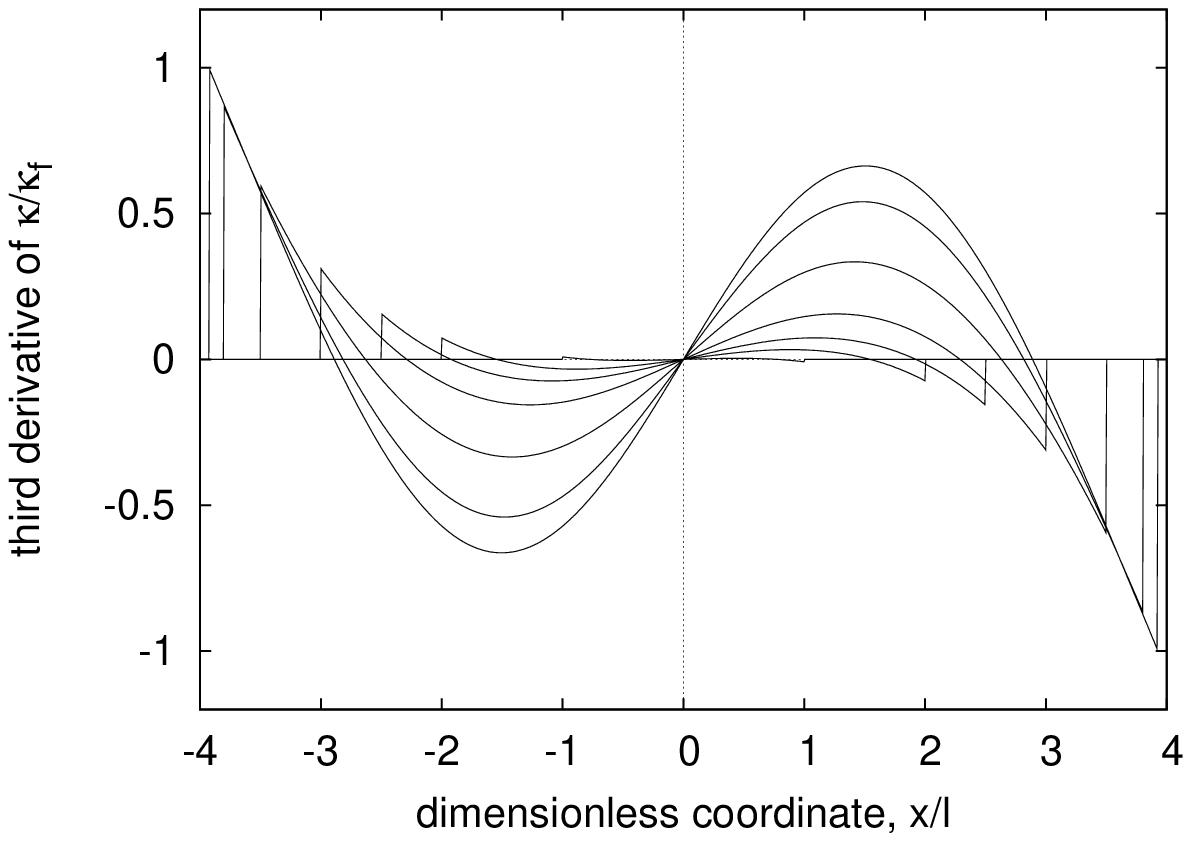}%
\end{tabular}
\caption{Explicit fourth-order gradient model, quadratic stress distribution: Evolution of (a) plastic strain,
(b) third derivative of plastic strain}%
\label{fig:aif4A_kap}%
\end{figure}

According to (\ref{eq:4qtanfx}),
variable $\delta_2$ continuously grows from its initial value $0$ at $\lamp=0$ to its maximum value
$\delta_{2,\max{}}\approx 0.779$ at $\lamp\approx 2.365$, and then decreases and tends to minus infinity as 
$\lamp$ approaches 3.9266; see Fig.~\ref{fig:comp_dl}.
Expressing now the integration constants $C_1$ and $C_3$ from (\ref{eq:cond4q_2})--(\ref{eq:cond4q_3}) 
and substituting into the general solution (\ref{eqgp63}),
we construct the particular solution
\beq
\kapt(\xi)=1-\frac{\cos\xi}{2\cos\lamp}-\frac{\cosh\xi}{2\cosh\lamp}
+\phi\left[\frac{\xi^2}{\lamg^2}-1
+\left(1+\frac{2-\lamp^2}{\lamg^2}\right)\frac{\cos\xi}{2\cos\lamp}+\left(1-\frac{2+\lamp^2}{\lamg^2}\right)\frac{\cosh\xi}{2\cosh\lamp}\right]
\eeq
The evolution of the plastic strain profile, graphically presented in Fig.~\ref{fig:aif4A_kap}a for $\lamg=4$,
is again monotonic.
The third derivative of plastic strain,
\beq
\kapt'''(\xi) = -\frac{\sin\xi}{2\cos\lamp}-\frac{\sinh\xi}{2\cosh\lamp}
+\phi\left[\left(1+\frac{2-\lamp^2}{\lamg^2}\right)\frac{\sin\xi}{2\cos\lamp}+\left(1-\frac{2+\lamp^2}{\lamg^2}\right)\frac{\sinh\xi}{2\cosh\lamp}\right]
\eeq
is discontinuous at the boundary of plastic zone, i.e., at points $\xi=\pm\lamp$.
This is documented by the graphs in Fig.~\ref{fig:aif4A_kap}b, plotted for $\lamg=4$.
The jumps in the third derivative are always positive, because $\kapt'''(-\lamp)>0$
and $\kapt'''(\lamp)<0$ for each admissible plastic zone size $\lamp$, and so the solution
is plastically admissible.

\begin{figure}
\centering
\begin{tabular}{cc}
(a) & (b) 
\\
\includegraphics[scale=0.7]{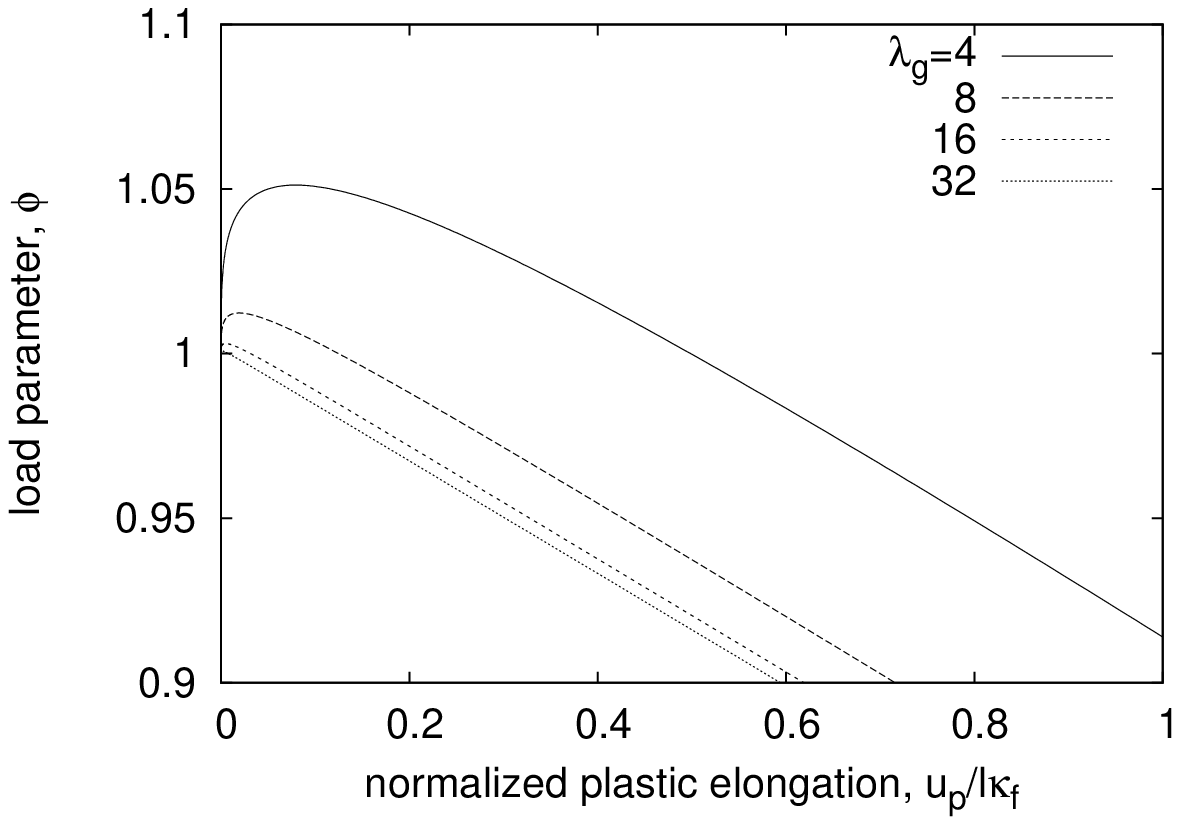}
&
\includegraphics[scale=0.7]{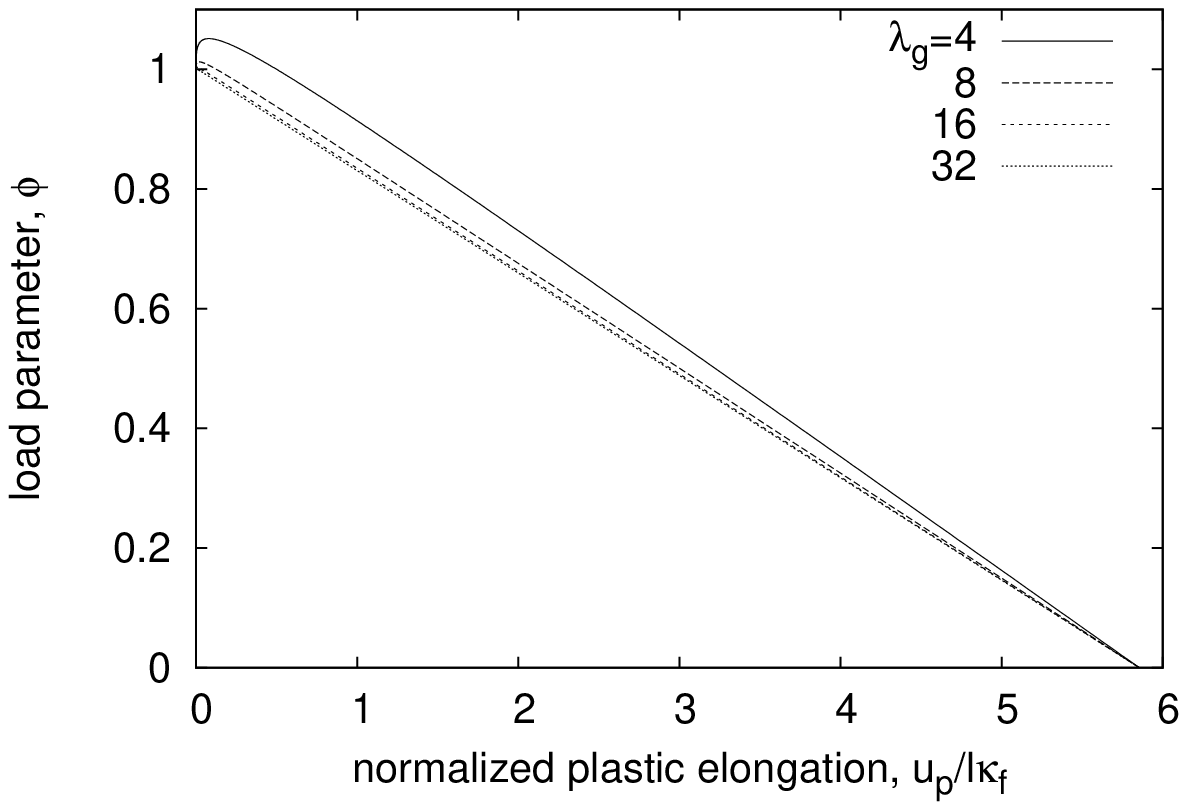}
\end{tabular}
\caption{Explicit fourth-order gradient model, quadratic stress distribution: Plastic part of load-displacement diagram for  different
values of $\lamg$ --- (a) early stages, (b) complete diagram}
\label{fig:aif4A_sd}%
\end{figure}

Integrating the normalized plastic strain along the plastic zone, we obtain the normalized plastic elongation
\begin{equation}
\frac{u_p}{l\kap_f}=\int_{-\lamp}^{\lamp} \kapt(\xi)\,\mbox{d}\xi =
\frac{\phi}{\lamg^2}\left[\frac{8(\tan\lamp-\lamp)(\tanh\lamp-\lamp)}{\tanh\lamp-\tan\lamp}-\frac{4\lamp^3}{3}\right]
\end{equation}
The resulting plastic part of the dimensionless load-displacement diagram is plotted in
Fig.~\ref{fig:aif4A_sd} for several values of parameter $\lamg$.

\subsection{Piecewise linear stress distribution}

For the piecewise linear stress distribution (\ref{eqgp9}) and the  explicit fourth-order model, 
the yield condition leads to a differential equation that can be written in the dimensionless form as
\begin{equation}
\kapt(\xi)-\kapt^{IV}(\xi)=1-\phi+\frac{\phi\vert\xi\vert}{\lamg}
\end{equation}
and has the general solution
\begin{equation}
\kapt(\xi)=\left\{\begin{array}{ll}
1-\phi-\phi\lamg^{-1}\xi
+C_1\cos{\xi}+C_2\sin{\xi}+C_3\cosh\xi+C_4\sinh\xi & \mbox{ for } -\lamp\le \xi \le 0
\\[2mm]
1-\phi+\phi\lamg^{-1}\xi
+C_5\cos{\xi}+C_6\sin{\xi}+C_7\cosh\xi+C_8\sinh\xi & \mbox{ for } 0\le \xi \le \lamp
\end{array}\right.
\label{eqgp86}
\end{equation}
If symmetry is ignored, eight integration constants plus two coordinates of the boundary of the plastic zone
can be determined from four continuity conditions at $\xi=0$
and three continuity conditions at each of the boundary points. 
By exploiting symmetry, we can reduce the
number of unknowns without affecting the solution. 
Symmetry conditions imply that 
\beq\label{eqgp110}
C_1=C_5,\;\; C_2=-C_6,\;\; C_3=C_7,\;\; C_4=-C_8
\eeq
and they lead to an automatic satisfaction
of continuity of $\kapt$ and $\kapt''$ at $\xi=0$. We still need to enforce  continuity of $\kapt'$ and $\kapt'''$ at $\xi=0$.
From these conditions combined with symmetry we obtain
\beq\label{eqgp110x}
C_6=C_8= -\frac{\phi}{2\lamg}
\eeq 
The remaining three unknowns, $C_5$, $C_7$ and $\lamp$, need to be determined from the conditions of twofold
continuous differentiability at $\xi=\lamp$. Under the assumption of monotonic expansion of the plastic zone,
these condition can be written in the same form (\ref{eqgp64}) as for the previous case with quadratic stress
distribution. After substitution from (\ref{eqgp86}) and (\ref{eqgp110x}) into  (\ref{eqgp64}), we obtain
a set of three equations
\bea
C_5\cos\lamp+C_7\cosh\lamp&=&\phi-1 +\frac{\phi}{2\lamg} (\sin\lamp+\sinh\lamp-2\lamp)
\\
-C_5\sin\lamp+C_7\sinh\lamp&=&\frac{\phi}{2\lamg} (\cos\lamp+\cosh\lamp-2)
\\
-C_5\cos\lamp+C_7\cosh\lamp&=&\frac{\phi}{2\lamg} (-\sin\lamp+\sinh\lamp)
\eea
and finally, after elimination of $C_5$ and $C_7$, we express the load parameter
in the form
\beq
\phi=\frac{\lamg}{\lamg-\delta_1}
\eeq
where the auxiliary parameter
\beq
\delta_1=\lamp+\frac{2-\cos^{-1}\lamp-\cosh^{-1}\lamp}{\tan\lamp-\tanh\lamp}
\eeq
depends only on parameter $\lamp$.
The relation between $\delta_1$ and $\lamp$ is shown graphically by the dashed curve in Fig.~\ref{fig:comp_dl}.

\begin{figure}
\centering
\begin{tabular}{cc}
(a) & (b) 
\\
\includegraphics[scale=0.7]{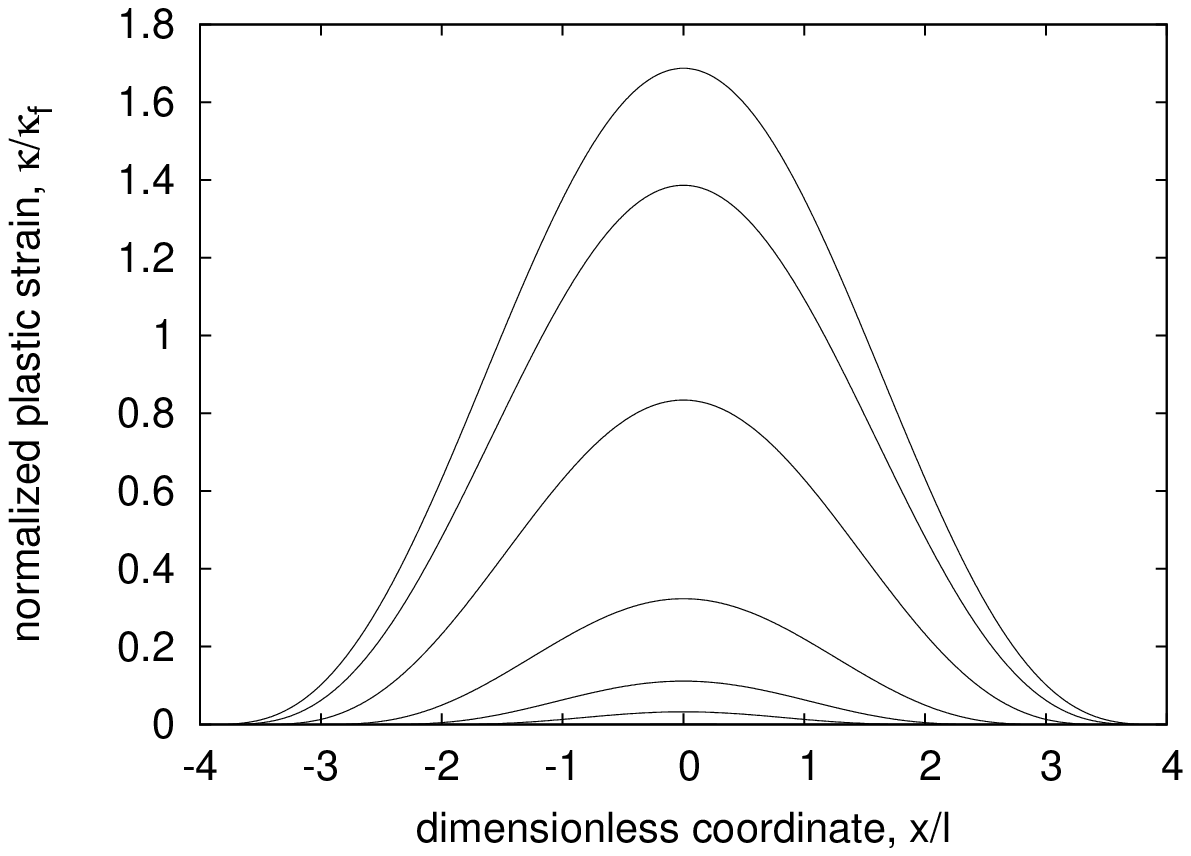}%
&
\includegraphics[scale=0.7]{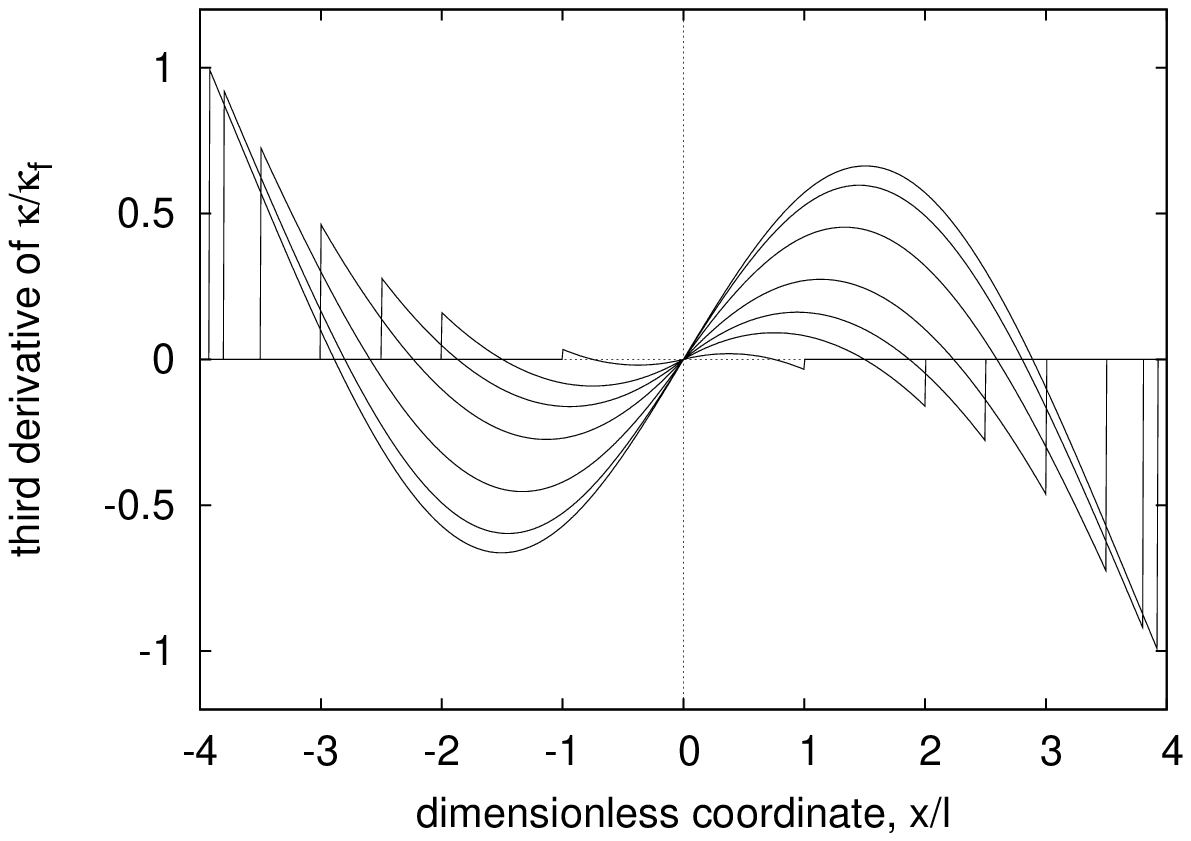}%
\end{tabular}
\caption{Explicit fourth-order gradient model, piecewise linear stress distribution: Evolution of (a) 
plastic strain, (b) third derivative of plastic strain}%
\label{fig:aif4B_kap}%
\end{figure}

After evaluation of the integration constants and their substitution into (\ref{eqgp86}), we obtain
the particular solution
\beql{eg:egp111}
\kapt(\xi) = \frac{\phi}{2\lamg}
\left[2\vert\xi\vert-2\delta_1+(\delta_1-\lamp+\sin\lamp)\frac{\cos\xi}{\cos\lamp}
+(\delta_1-\lamp+\sinh\lamp)\frac{\cosh\xi}{\cosh\lamp}-\sin\vert\xi\vert-\sinh\vert\xi\vert\right]
\eeq
The plots in Fig.~\ref{fig:aif4B_kap}a show a typical evolution of plastic strain, obtained
for  $\lamg=4$. The plastic zone again starts from the weakest section and expands monotonically, first at increasing and later
at decreasing axial force.
The third derivative of plastic strain,
\beq
\kapt'''(\xi) = \frac{\phi}{2\lamg}
\left[(\delta_1-\lamp+\sin\lamp)\frac{\sin\xi}{\cos\lamp}
+(\delta_1-\lamp+\sinh\lamp)\frac{\sinh\xi}{\cosh\lamp}+\sgn\xi(\cos\xi-\cosh\xi)\right]
\eeq
is discontinuous at the boundary of plastic zone, i.e., at points $\xi=\pm\lamp$.
This is documented by the graphs in Fig.~\ref{fig:aif4B_kap}b, plotted for $\lamg=4$.
Same as in the case of quadratic stress distribution, the jumps in the third derivative
are always positive, and the plastic admissibility of the solution is verified.

\begin{figure}
\centering
\begin{tabular}{cc}
(a) & (b) 
\\
\includegraphics[scale=0.7]{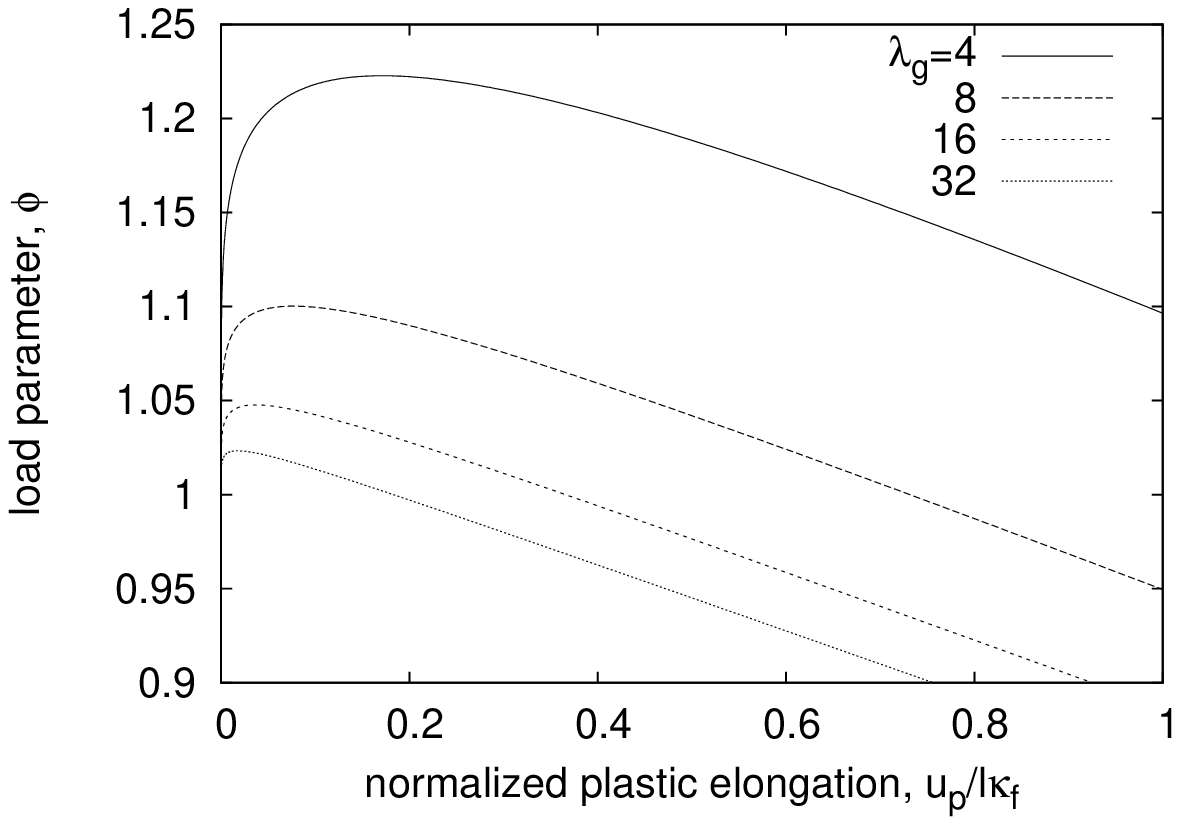}
&
\includegraphics[scale=0.7]{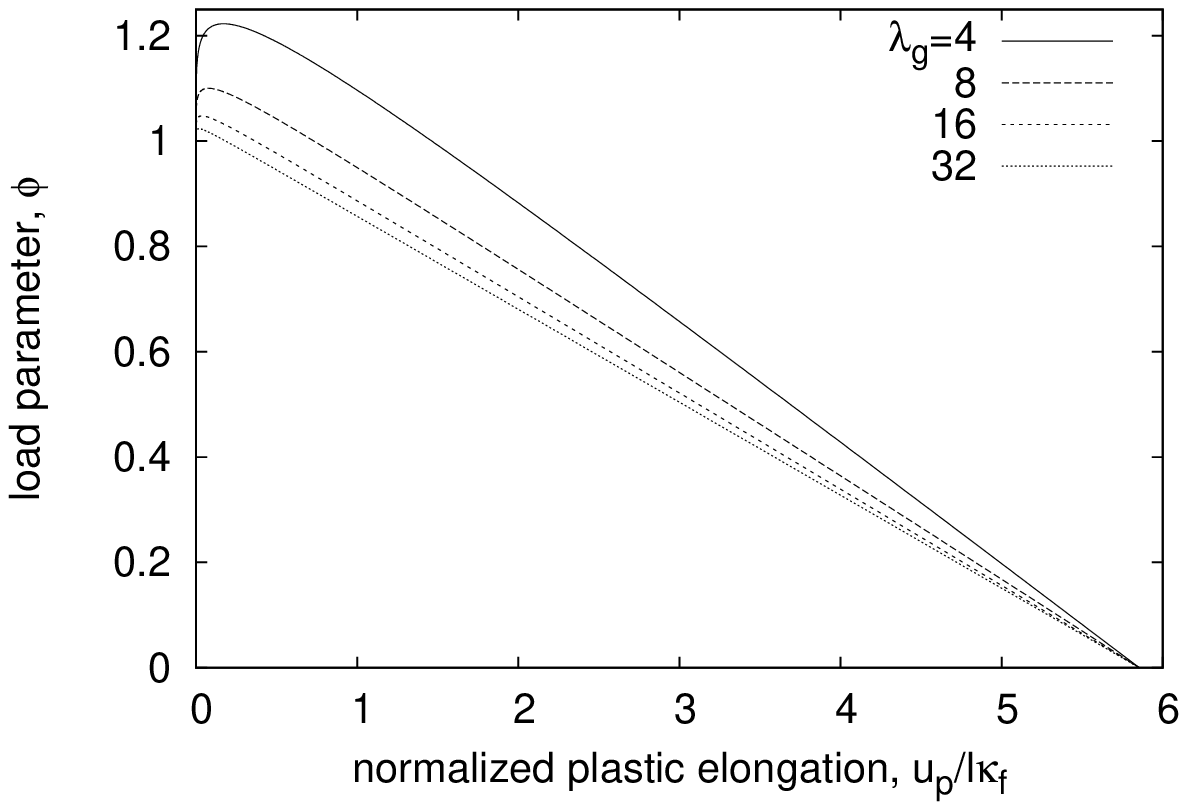}
\end{tabular}
\caption{Explicit fourth-order gradient model, piecewise linear stress distribution: Plastic part of load-displacement diagram for different
values of $\lamg$ --- (a) early stages, (b) complete diagram}
\label{fig:aif4B_sd}%
\end{figure}

Integrating the plastic strain (\ref{eg:egp111}), we obtain the plastic elongation,
which can be presented in the dimensionless form as
\beq
\frac{u_p}{l\kap_f}=
\frac{\phi}{\lamg}\left[\lamp^2-2\delta_1\lamp+(\delta_1-\lamp)(\tan\lamp+\tanh\lamp)+\cos^{-1}\lamp-\cosh^{-1}\lamp\right]
\eeq
The plastic part of the load-displacement diagram is depicted in Fig.~\ref{fig:aif4B_sd}
for several values of parameter $\lamg$. 

\section{Implicit gradient plasticity model}

The implicit gradient approach incorporates the influence of the
nonlocal cumulative plastic strain $\kapb$, defined as the solution
of the differential equation
\beq
\kapb(x)-l^2\bar\kappa''(x) = \kappa(x)
\label{eq:IGA2}
\eeq
usually with homogeneous Neumann boundary conditions
\beql{eq:igp21}
\kapb'=0 \hskip 10mm \mbox{on }\partial{\cal L}
\eeq
where ${\cal L}$ denotes the interval representing the bar and $\partial{\cal L}$
is its boundary, consisting of two points.
 In multiple dimensions,
the second spatial derivative is replaced by the Laplacean. 
The reason why $\kapb$ is called nonlocal is that its value 
at a given point $x$ depends on the values of the ``local''
cumulative plastic strain $\kap$ at all points of the body.
It can even be shown that $\kapb$ corresponds to
the weighted spatial average of $\kap$ with a special choice
of the weight function,
set equal to the Green function
of boundary value problem (\ref{eq:IGA2})--(\ref{eq:igp21}).

The simplest formulation of an implicit gradient plasticity model
could be based on the replacement of the local cumulative plastic
strain $\kap$ in the softening law by its nonlocal counterpart $\kapb$.
This formulation enforces nonzero dissipation but does not
prevent localization of plastic strain into a set of zero measure
(into a single cross section). In analogy to integral-type
nonlocal plasticity \cite{VerBri94,Stromberg96},
a finite size of the process zone can be obtained with $\kap$ replaced
by an ``overnonlocal'' variable $\kaph$, defined as the linear combination
\beq
\kaph = m\kapb+(1-m)\kap
\eeq
where parameter $m$ is larger than 1. In such a case, the linear softening
law is written as
\beq
 \sigma_Y = \sigma_0 + H\left[ m\bar\kappa + (1-m)\kappa\right]
\eeq
In the plastic zone, the yield condition $\sig=\sigY$ leads to the differential
equation
\beql{eq:igp1}
m\kapb(x) + (1-m)\kappa(x)=\frac{\sig(x)-\sig_0}{H}
\eeq
with two unknown functions, $\kapb$ and $\kap$, which are linked by 
(\ref{eq:IGA2}). Instead of solving a set of two differential equations,
it is convenient to substitute (\ref{eq:IGA2}) into  (\ref{eq:igp1})
and eliminate $\kap$. This leads to the equation
\beql{eq:igp2}
\bar\kap(x)+(m-1)l^2\bar\kap''(x)=\frac{\sig(x)-\sig_0}{H}
\eeq
with only one unknown function, $\kapb$. 

In terms of the
normalized functions $\kapt(\xi)=\kap(l\xi)/\kap_f$ and 
$\kaptb(\xi)=\kapb(l\xi)/\kap_f$, with $\kap_f=-\sig_0/H$,
equations (\ref{eq:IGA2}) and (\ref{eq:igp2}) are rewritten as
\bea\label{eq:igp3minus}
\kaptb(x)-\kaptb''(x) &=& \kapt(x)
\\
\label{eq:igp3}
\kaptb(\xi)+\mu^2\kaptb''(\xi)&=&1-\frac{\sig(\xi l)}{\sig_0}
\eea
where the prime denotes differentiation with respect to the dimensionless
coordinate $\xi=x/l$, and $\mu=\sqrt{m-1}$ is introduced for convenience.
To get the simplest possible form of equations, symbols $m$ and $\mu$ will
be used simultaneously, but they refer to a single independent material
parameter, and $m$ can always be replaced by $1+\mu^2$ or $\mu$ by $\sqrt{m-1}$.

Equation (\ref{eq:igp3}) is valid only in the plastic
zone, characterized by a positive value of the local variable $\kapt$.
Outside the plastic zone, we have $\kapt=0$ and $\kaptb$ is governed by
the homogeneous version of
equation (\ref{eq:igp3minus}), i.e., by
\beql{eq:igp4}
\kaptb(\xi)-\kaptb''(\xi)=0
\eeq

\subsection{Quadratic stress distribution}

For the quadratic stress distribution given by (\ref{eqgp7}), equation
(\ref{eq:igp3}) has the specific form
\begin{equation}
\kaptb(\xi)+\mu^2\kaptb''(\xi)=1-\phi+\frac{\phi\xi^2}{\lamg^{2}}
\label{eq:IGA5}
\end{equation}
where $\phi=\sigc/\sig_0$ is the load parameter.
The general solution in terms of the nonlocal variable 
\beq
\kaptb(\xi)=1-\phi\left(1+\frac{2\mu^2-\xi^2}{\lamg^{2}}\right)
+C_1\cos\frac{\xi}{\mu}+C_2\sin\frac{\xi}{\mu}
\eeq
substituted into  (\ref{eq:igp3minus})
gives the local variable
\beql{eq:igp8}
\kapt(\xi)=\kaptb(\xi)-\kaptb''(\xi)=
1-\phi\left(1+\frac{2m-\xi^2}{\lamg^{2}}\right)
+\frac{m}{\mu^2}\left(C_1\cos\frac{\xi}{\mu}+C_2\sin\frac{\xi}{\mu}\right)
\eeq
By virtue of symmetry we get $C_2=0$.

Equations (\ref{eq:IGA5})--(\ref{eq:igp8}) are valid only in the plastic zone $\Ip$ characterized
by $\kapt>0$. Outside the plastic zone, the local variable $\kapt$ vanishes
and the nonlocal variable $\kaptb$ is governed
by the homogeneous differential equation (\ref{eq:igp4}) with general
solution
\beq
\kaptb(\xi) = C_3{\mbox e}^{\xi}+C_4{\mbox e}^{-\xi}
\eeq
At the physical boundary, the homogeneous Neumann boundary condition $\kaptb'=0$
is usually imposed. For a finite bar, the solution then depends
on the bar length as an additional 
parameter. For simplicity, we assume that the bar is much longer than the
plastic zone. The boundary condition is then imposed ``at infinity'',
which means that it is replaced
by the requirement that the solution must remain bounded.
In that case, in the ``right'' part of the elastic zone integration
constant $C_3$ must vanish.
Integration constants $C_1$ and $C_4$
and the dimensionless plastic zone size $\lamp$ can be calculated 
from the condition $\kapt(\lamp)=0$ and from the conditions of continuous
differentiability of $\kaptb$ at $\xi=\lamp$.

After elimination of $C_1$
and $C_4$ we end up with a single equation that links the plastic zone size
$\lamp$ and the load parameter $\phi$, of course with an influence of parameters
$\lamg$ and $m$. It turns out that, similar to (\ref{eq:phi1}) and (\ref{eq58}), 
the load parameter can be expressed in the form
\beql{eq:igp5}
\phi=\frac{\lamg^2}{\lamg^2-\delta_2}
\eeq
but the auxiliary parameter $\delta_2$ is instead of (\ref{eqgpdelta2}) or (\ref{eq:4qtanfx}) given by
\beql{eq:igp6}
\delta_2 = \lamp^2 +2m \frac{\lamp-\mu\tan\frac\lamp\mu}{1+\mu\tan\frac\lamp\mu}
\eeq
\begin{figure}
\centering
\includegraphics[scale=0.7]{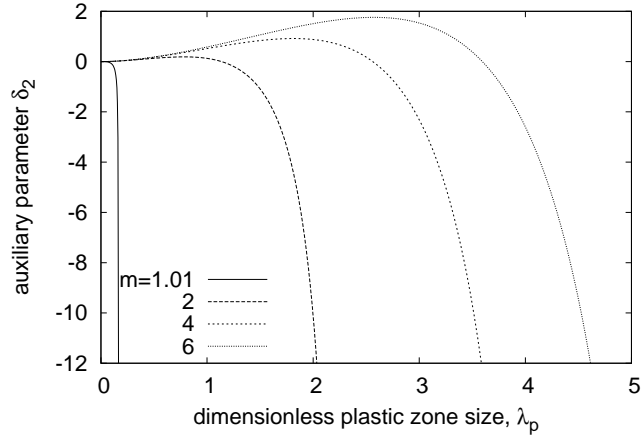}
\caption{Implicit gradient model, quadratic stress distribution: Dependence between auxiliary parameter $\delta_2$ and dimensionless plastic zone size $\lamp$}
\label{fig:ImpA-dl}%
\end{figure}
The dependence of $\delta_2$ on $\lamp$ is graphically illustrated in
Fig.~\ref{fig:ImpA-dl} for several values of parameter $m$.
Once again, the advantage of using the auxiliary parameter is that parameter $\lamg$ appears only in (\ref{eq:igp5})
and parameters $\lamp$ and $m$ or $\mu$ only in (\ref{eq:igp6}).
The integration constant is then evaluated as
\beql{eq:igp7}
C_1 = \frac{2\phi}{\lamg^2}\,\frac{(1+\lamp)\mu^2}{\cos\frac\lamp\mu+\mu\sin\frac\lamp\mu}
\eeq
and
the distribution of plastic strain in the process zone 
\beql{eq:igp8x}
\kapt(\xi) = \frac{\phi}{\lamg^2}
\left[\xi^2-\lamp^2-\frac{2m(1+\lamp)}{1+\mu\tan\frac\lamp\mu}
\left(1-
\frac{\cos\frac\xi\mu}{\cos\frac\lamp\mu}\right)\right]
\eeq
is obtained by substituting (\ref{eq:igp7}) and $C_2=0$ into (\ref{eq:igp8}).

\begin{figure}
\centering
\begin{tabular}{cc}
(a) & (b) 
\\
\includegraphics[scale=0.7]{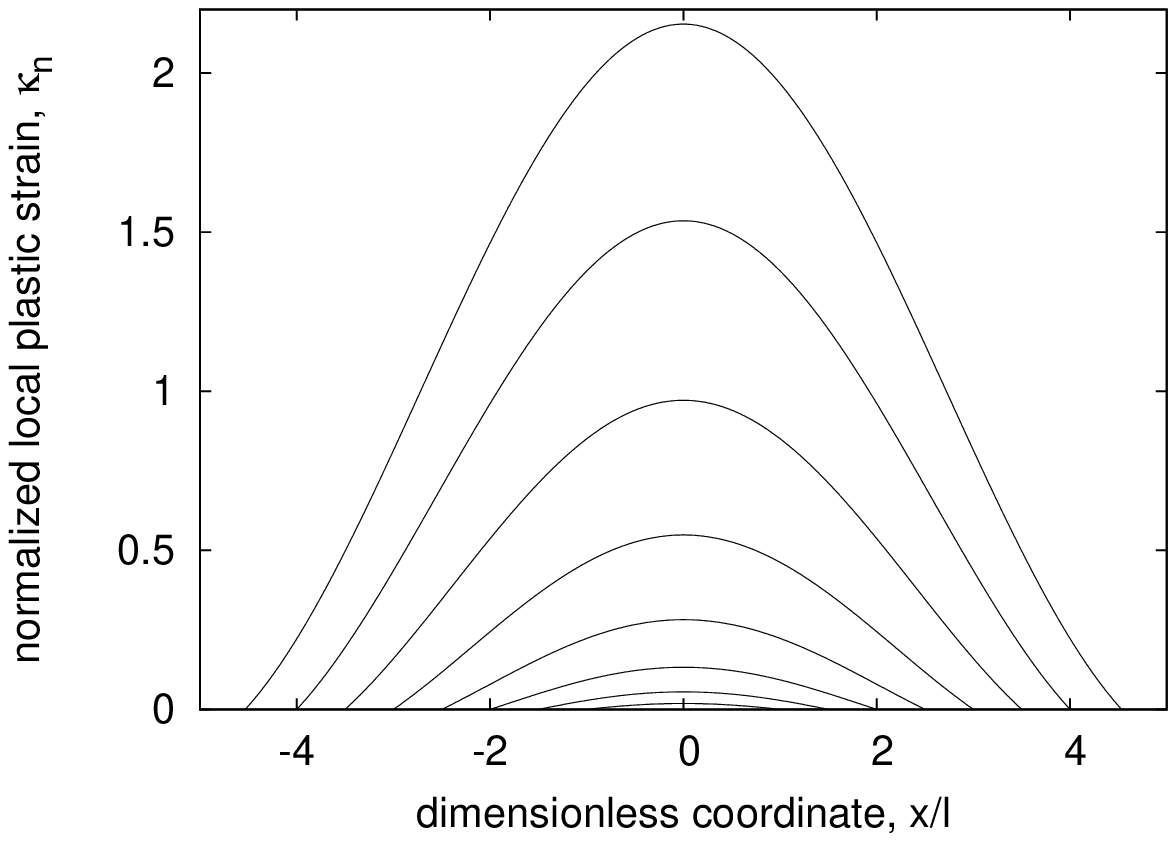}
&
\includegraphics[scale=0.7]{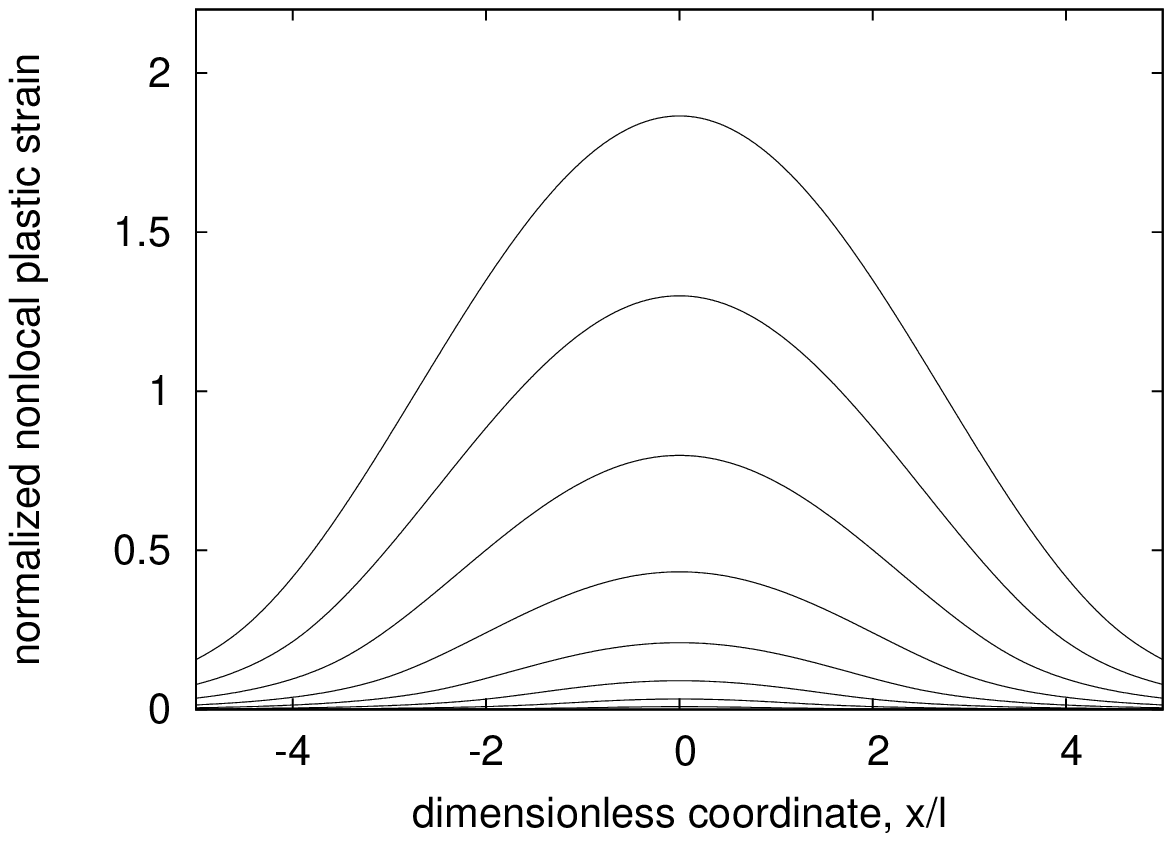}
\end{tabular}
\caption{Implicit gradient model, quadratic stress distribution: Evolution of normalized plastic strain profiles for $m=4$ and $\lamg=5$ --- (a) local plastic strain, (b) nonlocal plastic strain}
\label{fig:ImpA-kap}
\end{figure}

\begin{figure}
\centering
\includegraphics[scale=0.7]{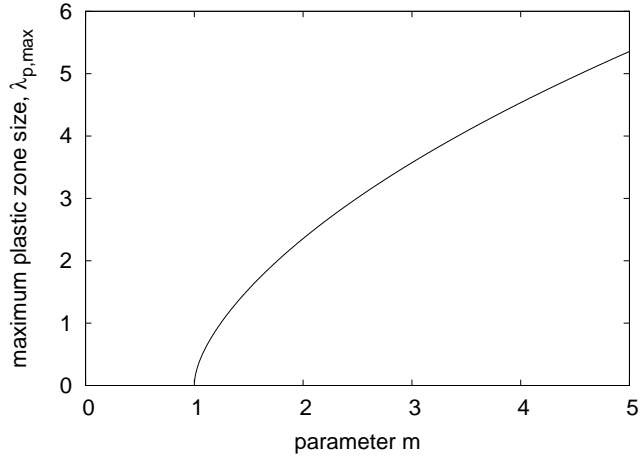}
\caption{Implicit gradient model: Dependence of the maximum dimensionless half-size of plastic zone, $\lambda_{p,\max{}}$, on model parameter $m$}
\label{fig:ImpA-lm}%
\end{figure}

An example of the evolution of normalized local and nonlocal plastic strains
for $m=4$ and $\lamg=5$ is presented in Fig.~\ref{fig:ImpA-kap}.
Note that the local plastic strain is continuous but not continuously
differentiable at the boundary of the plastic zone. In this aspect,
the implicit gradient model differs from the explicit one. Outside the
plastic zone, the local plastic strain vanishes but the nonlocal
plastic strain does not.
The evolution of local plastic strain is monotonic and the 
 plastic zone expands up to its maximum length 
\begin{equation}
2\lambda_{p,\max{}}=2\mu\left(\pi-\arctan\frac{1}{\mu}\right)
\label{eq:lampmax}
\end{equation}
which depends on parameter $\mu$ 
and tends to zero as $\mu$ approaches zero, i.e.,
as $m=1+\mu^2$  approaches 1 from the right; see Fig.~\ref{fig:ImpA-lm}. This confirms that the simple model with $m=1$,
i.e., with softening driven by the nonlocal cumulative plastic strain $\kapb$,
does not act as a genuine localization limiter.

\begin{figure}
\centering
\begin{tabular}{cc}
(a) & (b) 
\\
\includegraphics[scale=0.7]{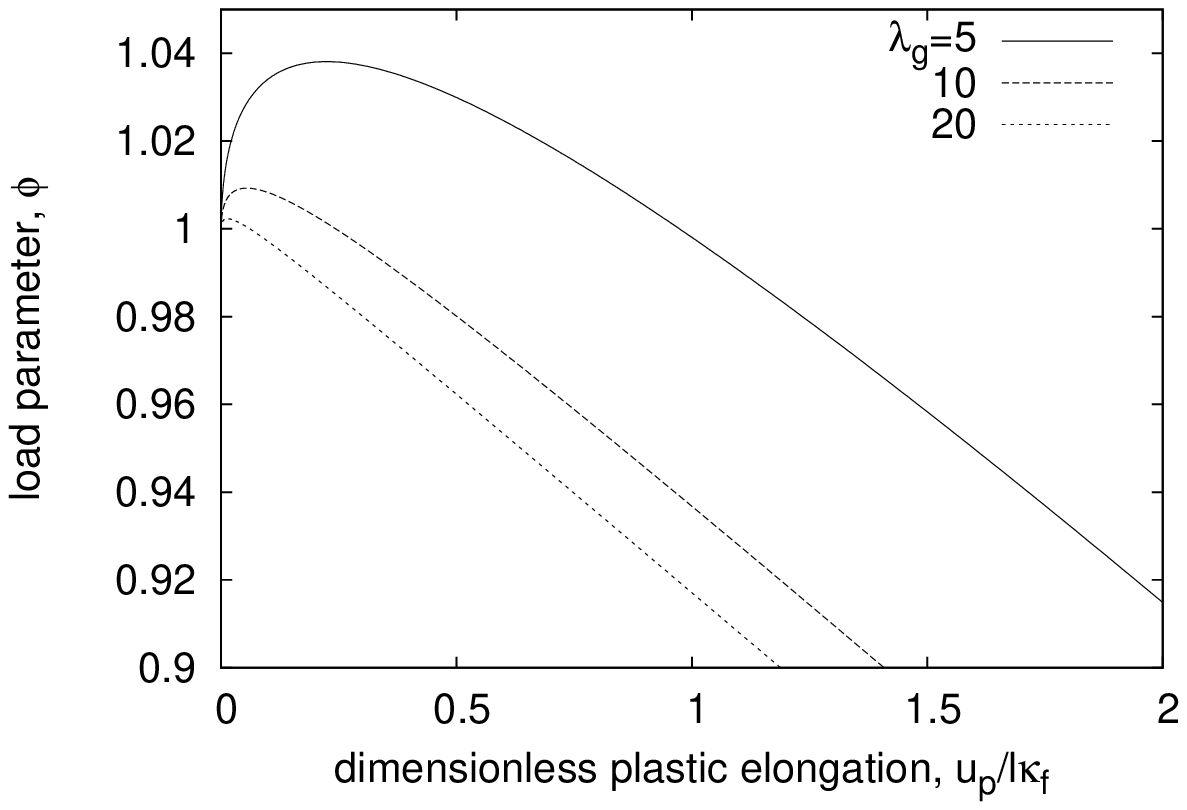}
&
\includegraphics[scale=0.7]{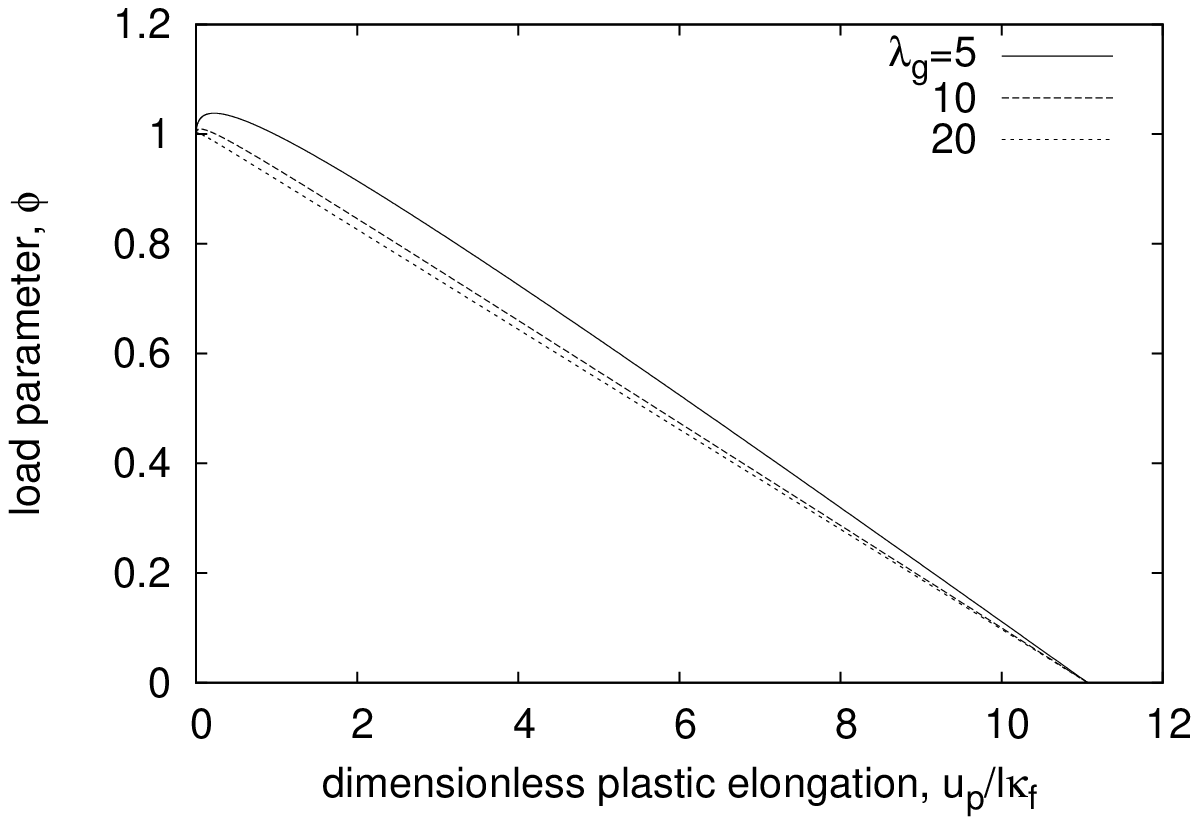}
\end{tabular}
\caption{Implicit gradient model, quadratic stress distribution: Plastic part of load-displacement diagram for $m=4$ and different
values of $\lamg$ --- (a) early stages, (b) complete diagram}
\label{fig:ImpA-sd_m2}
\end{figure}

Integration of the normalized plastic strain leads to the normalized plastic elongation
\beq
\frac{u_p}{l\kap_f}= \frac{2\phi}{\lamg^2}
\left[\frac{2m(1+\lamp)}{1+\mu\tan\frac\lamp\mu}\left(\mu\tan\frac\lamp\mu-\lamp\right)-\frac{2\lamp^3}{3}\right]
\eeq
The plastic part of the load-displacement diagram is shown in
Fig.~\ref{fig:ImpA-sd_m2} for different values of parameter $\lamg$.
For smaller values of $\lamg$, the peak load is higher and the dissipated
energy as well. For large values of $\lamg$, the diagram is close to a linear 
one.

\subsection{Piecewise linear stress distribution}

\begin{figure}
\centering
\includegraphics[scale=0.7]{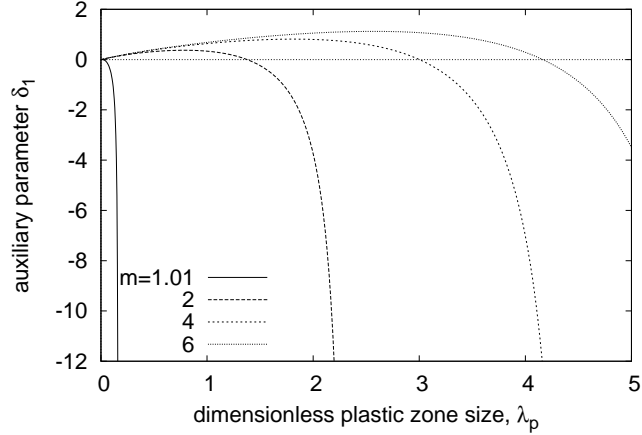}
\caption{Implicit gradient model, piecewise linear stress distribution: Dependence between auxiliary parameter $\delta_1$ and dimensionless plastic zone size $\lamp$}
\label{fig:ImpB-dl}%
\end{figure}

For piecewise linear stress distribution (\ref{eqgp9}), equation
(\ref{eq:igp3}) has the specific form
\begin{equation}
\kaptb(\xi)+\mu^2\kaptb''(\xi)=1-\phi+\frac{\phi\vert\xi\vert}{\lamg}
\label{eq:IGA5B}
\end{equation}
For $\xi\in[0,\lamp]$, the general solution reads
\beql{eq:igp22}
\kaptb(\xi) = 1-\phi+\frac{\phi\xi}{\lamg} +C_1\cos\frac{\xi}{\mu}+C_2\sin\frac{\xi}{\mu}
\eeq
and from symmetry and continuous differentiability we get the condition
$\kaptb'(0)=0$, which implies that
\beq
C_2 = - \frac{\mu\phi}{\lamg}
\eeq
At the boundary of the plastic zone, we impose the condition of vanishing local
plastic strain, $\kapt(\lamp)=0$,
which can be rewritten in terms of the nonlocal plastic strain as
$\kaptb(\lamp)=\kaptb''(\lamp)$. Continuous differentiability of the nonlocal
plastic strain combined with  boundedness in the semiinfinite
elastic zone leads to the condition $\kaptb(\lamp)=-\kaptb'(\lamp)$. 
In this way, two equations for two unknowns $C_1$ and $\lamp$ are constructed.
Eliminating the integration constant $C_1$, we arrive at one single equation
that links the load parameter $\phi$ and the plastic zone size $\lamp$. After some algebra,
the load parameter can be expressed  as
\beq
\phi = \frac{\lamg}{\lamg-\delta_1}
\eeq
where 
\beql{eq:igp23}
\delta_1 = \lamp-\frac{2m\sin^2\frac{\lamp}{2\mu}}{\cos\frac{\lamp}{\mu}+\mu\sin\frac{\lamp}{\mu}}
\eeq
is an auxiliary parameter that depends only on $\lamp$ and $m$ but not on $\lamg$.
The relation between $\delta_1$ and $\lamp$ described  by (\ref{eq:igp23}) 
is graphically presented in Fig.~\ref{fig:ImpB-dl} for several values of parameter
$m$.
The plastic zone expands to its maximum size, given again by formula
(\ref{eq:lampmax}), which was graphically presented in Fig.~\ref{fig:ImpA-lm}. 

\begin{figure}
\centering
\begin{tabular}{cc}
(a) & (b) 
\\
\includegraphics[scale=0.7]{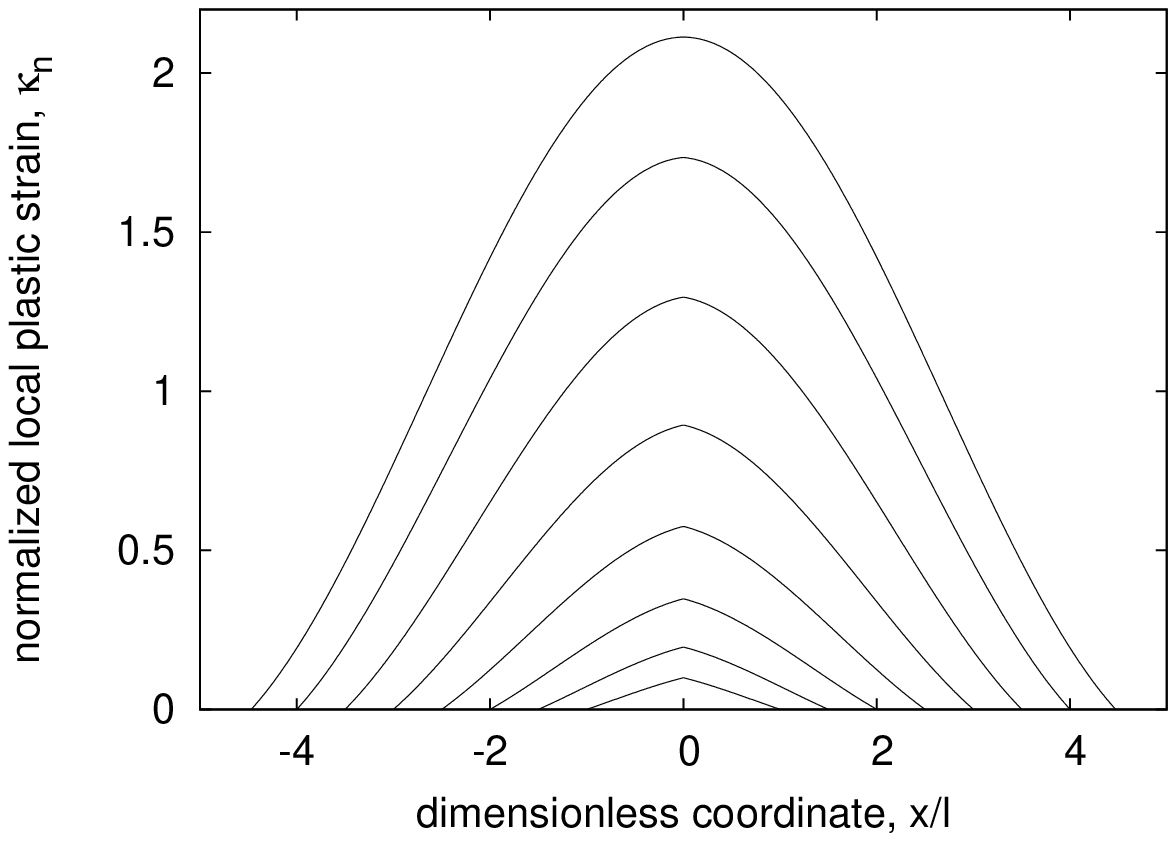}
&
\includegraphics[scale=0.7]{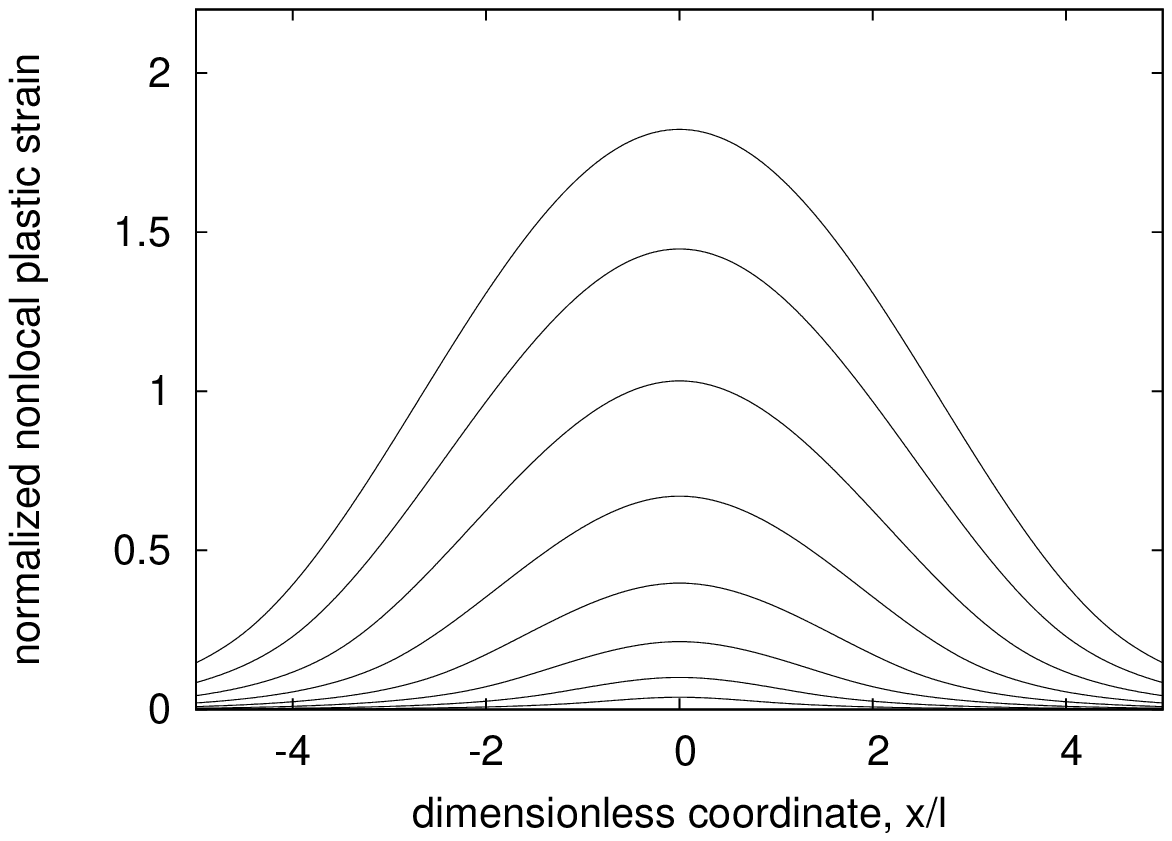}
\end{tabular}
\caption{Implicit gradient model, piecewise linear stress distribution: Evolution of normalized plastic strain profiles for $m=4$ and $\lamg=5$ --- (a) local plastic strain, (b) nonlocal plastic strain}
\label{fig:ImpB-kap}
\end{figure}

For each admissible size of plastic zone $\lamp$, integration constant $C_1$ 
is readily expressed,
and its substitution into (\ref{eq:igp22}) provides the particular solution. To make the 
resulting expression valid in the entire plastic zone, we replace $\xi$
by its absolute value, accounting for symmetry. The final formula for
the normalized nonlocal plastic strain
is then 
\beq
\kaptb(\xi) = \frac{\phi}{\lamg}\left[\vert\xi\vert-\delta_1
+\mu\left(\frac{1+\lamp-\delta_1-\mu\sin\frac{\lamp}{\mu}-\cos\frac{\lamp}{\mu}}{\sin\frac{\lamp}{\mu}-\mu\cos\frac{\lamp}{\mu}}\cos\frac{\xi}{\mu}
-\sin\frac{\vert\xi\vert}{\mu}\right)\right]
\eeq
and the corresponding normalized local plastic strain is given by
\bea\nonumber
\kapt(\xi)&=&\kaptb(\xi)-\kaptb''(\xi)=
\\ 
&=&
\frac{\phi}{\lamg}\left[\vert\xi\vert-\delta_1
+\frac{m}{\mu}\left(\frac{1+\lamp-\delta_1-\mu\sin\frac{\lamp}{\mu}-\cos\frac{\lamp}{\mu}}{\sin\frac{\lamp}{\mu}-\mu\cos\frac{\lamp}{\mu}}\cos\frac{\xi}{\mu}
-\sin\frac{\vert\xi\vert}{\mu}\right)\right]
\label{eq:igp24}
\eea
An example of evolution of the plastic strain profile is shown in Fig.~\ref{fig:ImpB-kap}.
Note the discontinuity in the derivative of the local plastic strain at $\xi=0$,
which is clearly manifested at early stages of evolution.

\begin{figure}
\centering
\begin{tabular}{cc}
(a) & (b) 
\\
\includegraphics[scale=0.7]{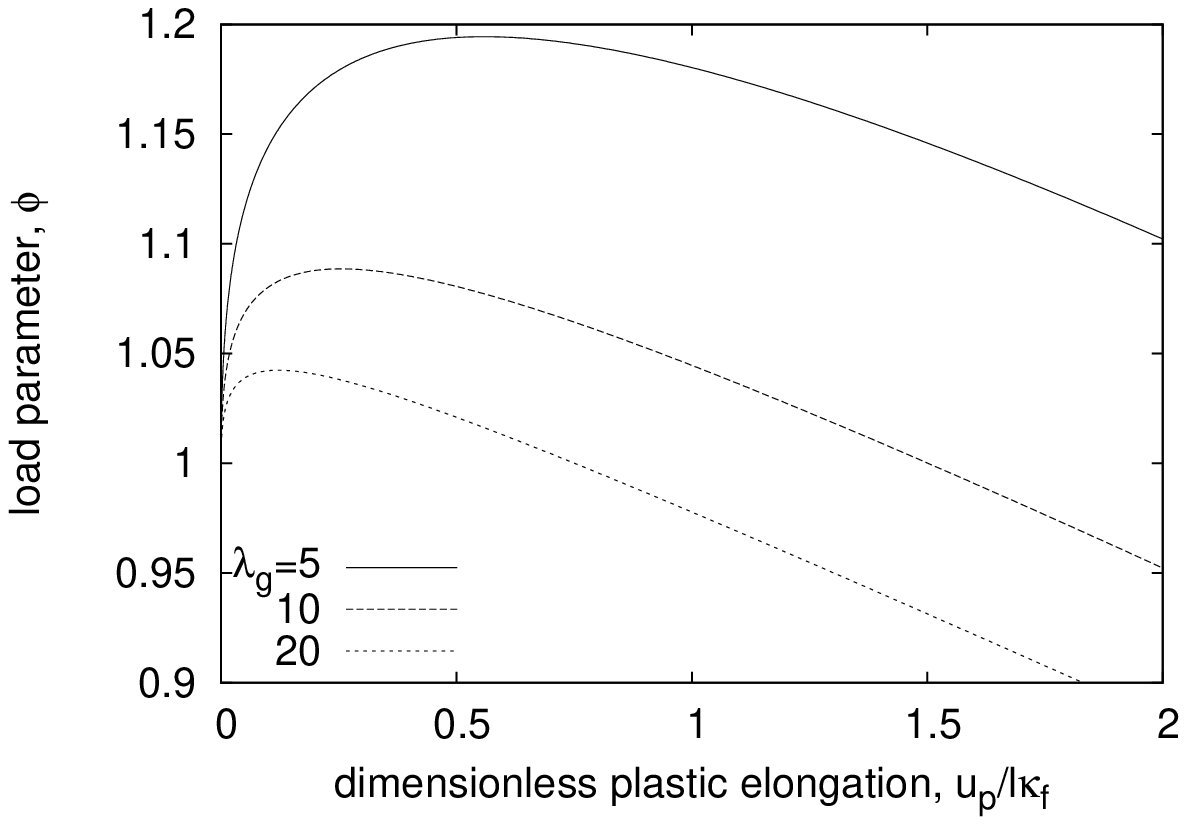}
&
\includegraphics[scale=0.7]{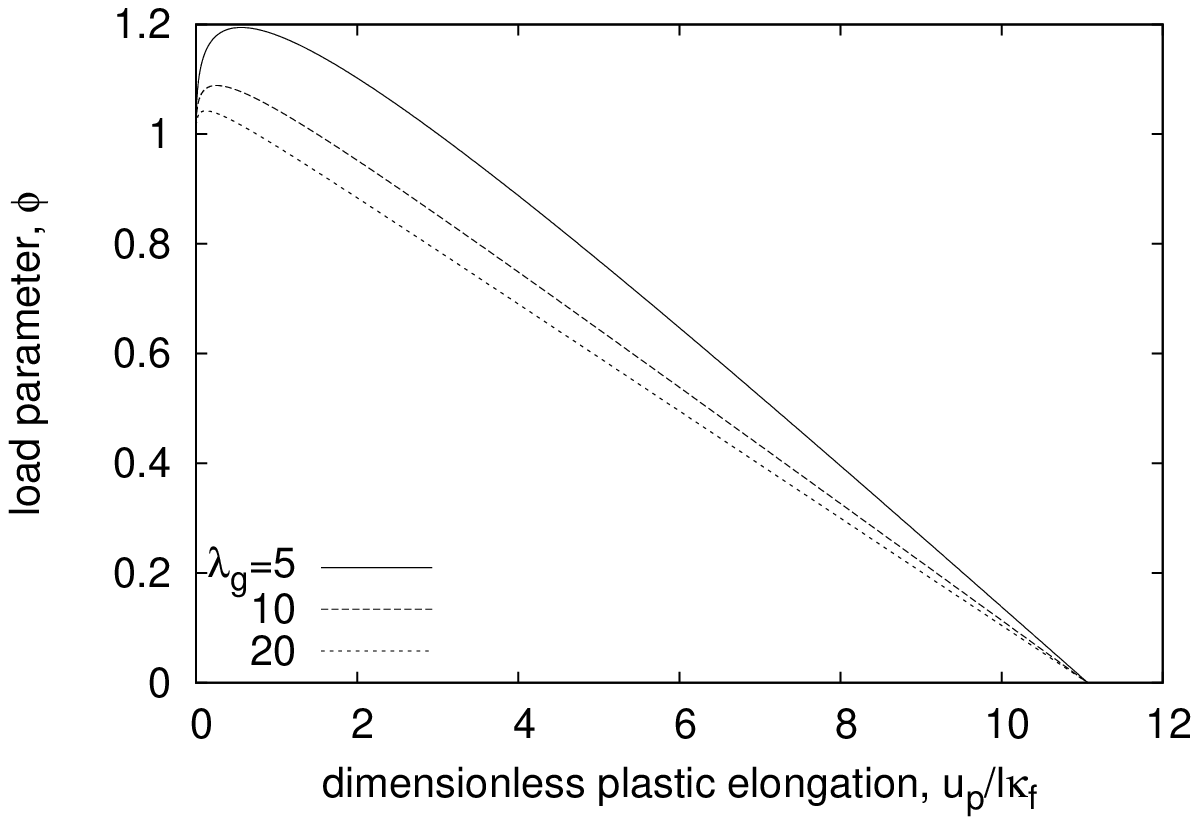}
\end{tabular}
\caption{Implicit gradient model, piecewise linear stress distribution:Plastic part of load-displacement diagram for $m=4$ and different
values of $\lamg$ --- (a) early stages, (b) complete diagram}
\label{fig:ImpB-sd_m2}
\end{figure}

Integrating  (\ref{eq:igp24})
over the plastic zone, we obtain the dimensionless plastic elongation
\beq
\frac{u_p}{l\kap_f}=\frac{\phi}{\lamg}
\left[\lamp(\lamp-2\delta_1)
-2m\frac{(\delta_1-\lamp)\sin\frac{\lamp}{\mu}+\mu(1-\cos\frac{\lamp}{\mu})}{\sin\frac{\lamp}{\mu}-\mu\cos\frac{\lamp}{\mu}}\right]
\eeq
The corresponding load-displacement diagram is plotted in Fig.~\ref{fig:ImpB-sd_m2}.

\section{Implicit gradient plasticity with modified boundary conditions}

The usual formulation of implicit gradient plasticity, developed
by Geers and coworkers \cite{GeeEngUba01,EngGeeBaa02,Gee04} and considered
in the previous section, 
imposes the homogeneous Neumann boundary 
condition at the physical boundary of the body of interest. 
In a recent study focusing on applications to beam bending,
Challamel \cite{Cha08} proposed to impose that condition on the boundary
of the plastic zone. For comparison, we will present the solution
of the one-dimensional localization problem based on this modification.
The only difference compared to the analysis
performed in the previous section is that the requirement of boundedness of the
nonlocal plastic strain is replaced by the conditions $\kaptb'(\lamp)=0$
and $\kaptb'(-\lamp)=0$. Since the analysis proceeds along the same line as
in the previous section, we omit all the intermediate steps
and directly proceed to the results and their discussion.

\subsection{Quadratic stress distribution}

The auxiliary parameter $\delta_2$ related to the load parameter $\phi$ by (\ref{eq:igp5}) is expressed as
\beq
\delta_2 = \lamp^2+2m\left(\frac{\lamp}{\mu}\,\cotan\frac{\lamp}{\mu}-1\right)
\eeq
and the distribution of nonlocal and local plastic strain
is given by 
\bea\label{eq103}
\kaptb(\xi) &=&
\left\{ \begin{array}{ll} 
\displaystyle\frac{\phi}{\lamg^2}\left(\xi^2-\delta_2-2\mu^2+\frac{2\mu\lamp}{\sin\frac{\lamp}{\mu}}\cos\frac{\xi}{\mu}\right) & \mbox{ for } \xi\in\Ip=(-\lamp,\lamp)
\\
\displaystyle\frac{2\phi}{\lamg^2}\left(1-\frac{\lamp}{\mu}\,\cotan\frac{\lamp}{\mu}\right)\cosh(\vert\xi\vert-\lamp)
 & \mbox{ for } \xi\in\Ie={\cal L}\setminus\Ip
\end{array}\right.
\\[3mm]
\kapt(\xi) &=&
\left\{ \begin{array}{ll} 
\displaystyle\frac{\phi}{\lamg^2}\left(\xi^2-\delta_2-2m+\frac{2m\lamp}{\mu\sin\frac{\lamp}{\mu}}\cos\frac{\xi}{\mu}\right) & \mbox{ for } \xi\in\Ip=(-\lamp,\lamp)
\\
0 & \mbox{ for } \xi\in\Ie={\cal L}\setminus\Ip
\end{array}\right.
\eea
The plastic zone expands up to its maximum size 
\beq
2\lambda_{p,\max{}}=2\pi\mu
\eeq
and the normalized plastic elongation is
\beq
\frac{u_p}{l\kap_f}=\frac{2\phi\lamp}{\lamg^2}
\left(\frac{\lamp^2}{3}-\delta_2\right)
\eeq

\begin{figure}
\centering
\begin{tabular}{cc}
(a) & (b) 
\\
\includegraphics[scale=0.7]{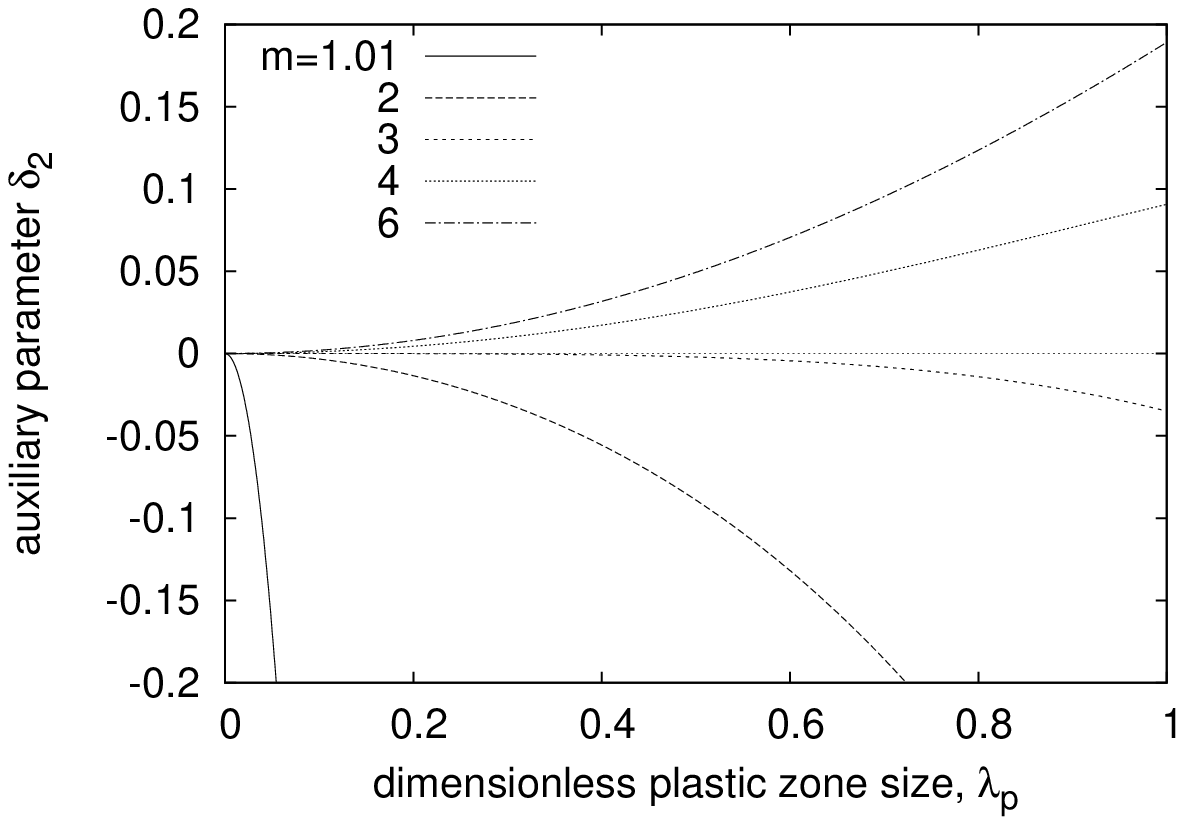}
&
\includegraphics[scale=0.7]{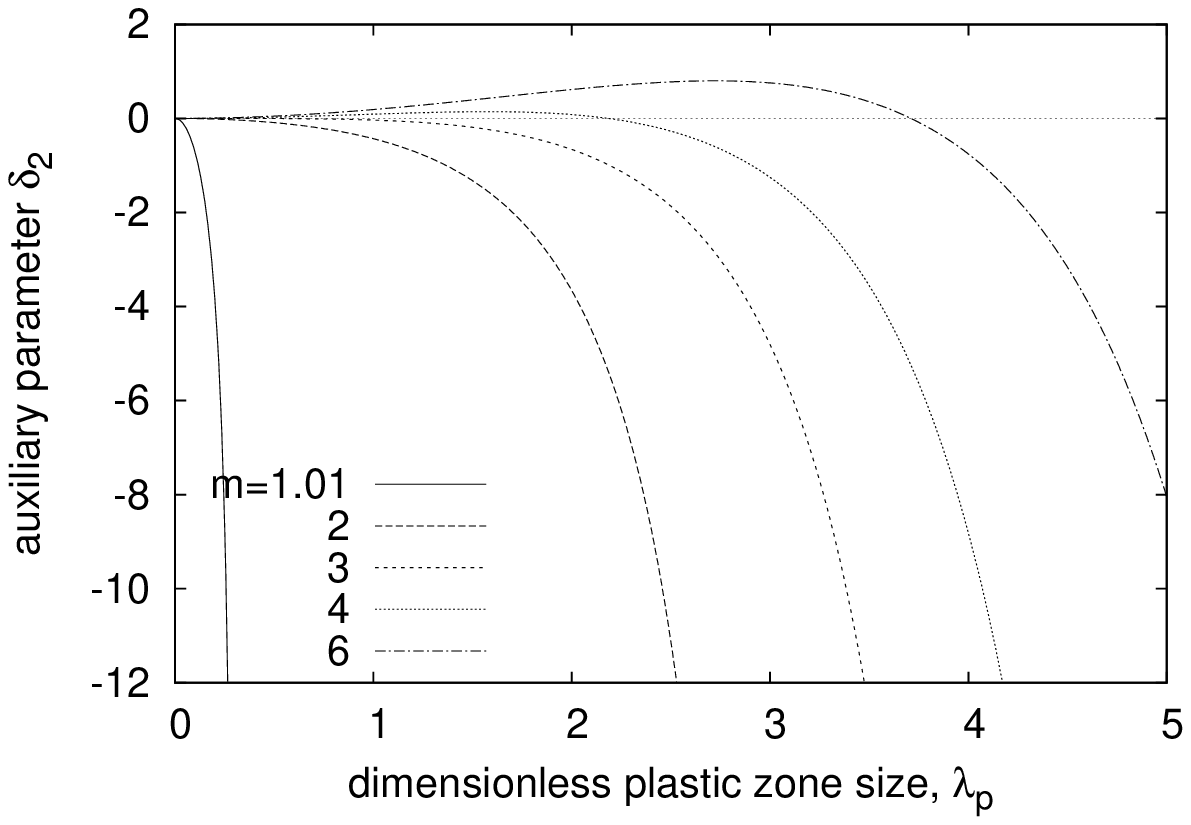}
\end{tabular}
\caption{Modified implicit gradient model, quadratic stress distribution: Dependence between auxiliary parameter $\delta_2$ and dimensionless plastic zone size $\lamp$ ---
(a) early stages, (b) global picture}
\label{fig:ImpA-dl2}%
\end{figure}

\begin{figure}
\centering
\begin{tabular}{cc}
(a) & (b) 
\\
\includegraphics[scale=0.7]{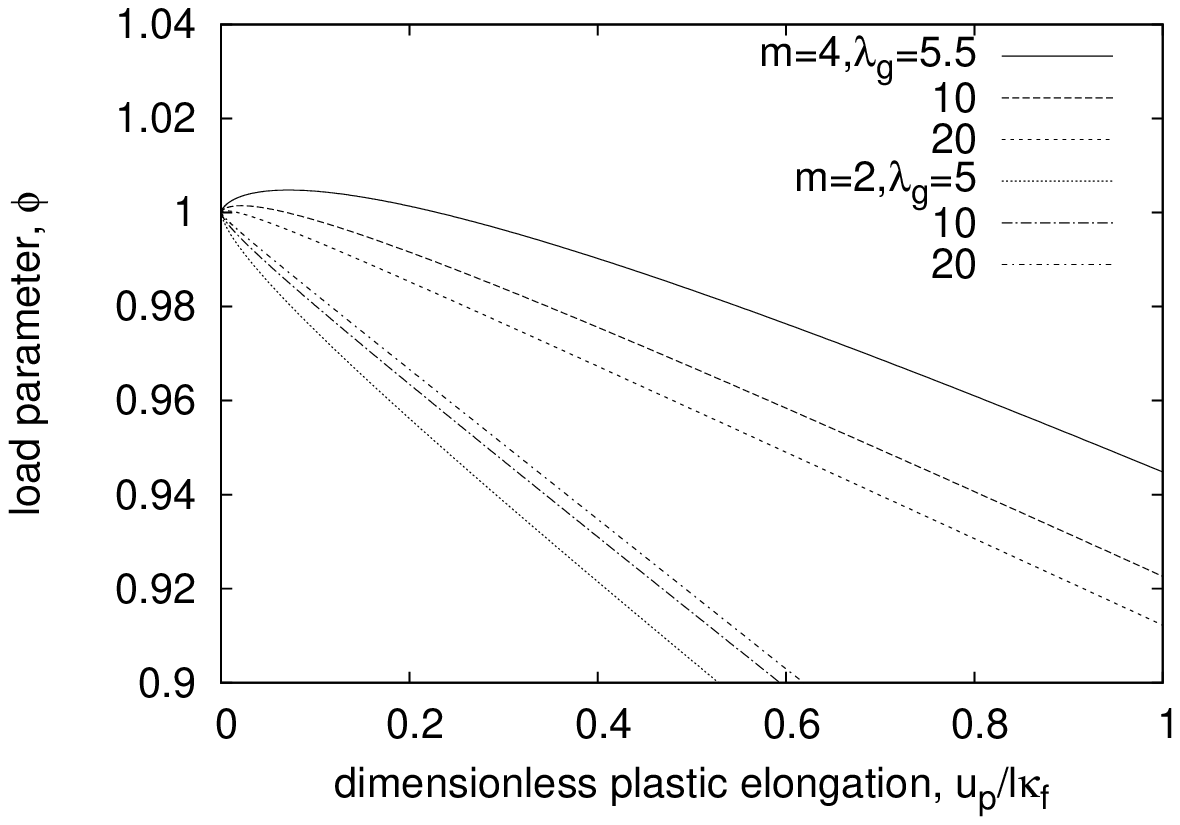}
&
\includegraphics[scale=0.7]{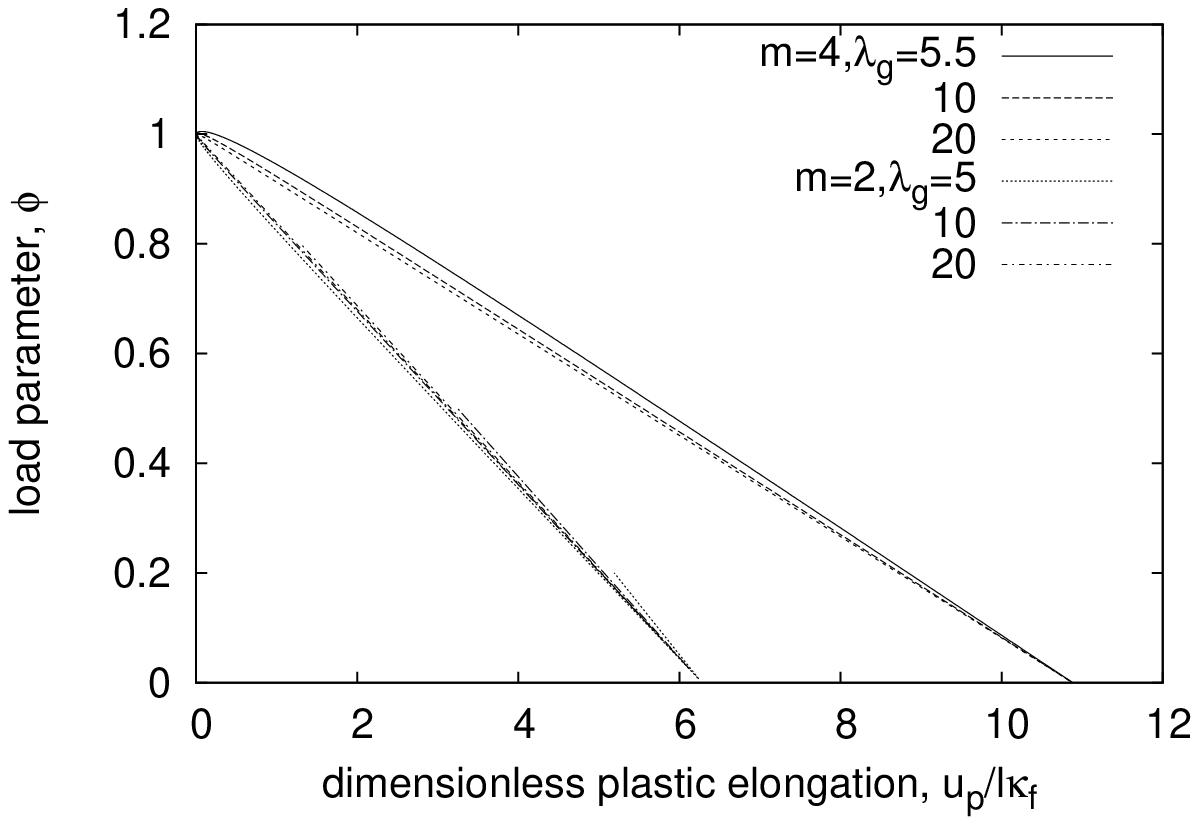}
\end{tabular}
\caption{Modified implicit gradient model, quadratic stress distribution: Plastic part of load-displacement diagram for different
values of parameters $m$ and $\lamg$ --- (a) early stages, (b) complete diagram}
\label{fig:ImpA2-sd_m4}
\end{figure}

The dependence of the auxiliary parameter $\delta_2$ on the plastic zone size is graphically
shown in Fig.~\ref{fig:ImpA-dl2} for several values of model parameter $m$. Interestingly,
if $m$ is in the range between 1 and 3, parameter $\delta$ monotonically decreases 
from its   
initial value 0 at $\lamp=0$ and never becomes positive. This means that the corresponding load parameter $\phi$
given by (\ref{eq:igp5})
monotonically decreases from its initial value 1, and the global response does not exhibit any hardening at all;
see Fig.~\ref{fig:ImpA2-sd_m4}.
In this aspect, the current model differs from all the other models covered by the present study.
Global hardening arises only for $m>3$. 

\begin{figure}
\centering
\begin{tabular}{cc}
(a) & (b) 
\\
\includegraphics[scale=0.7]{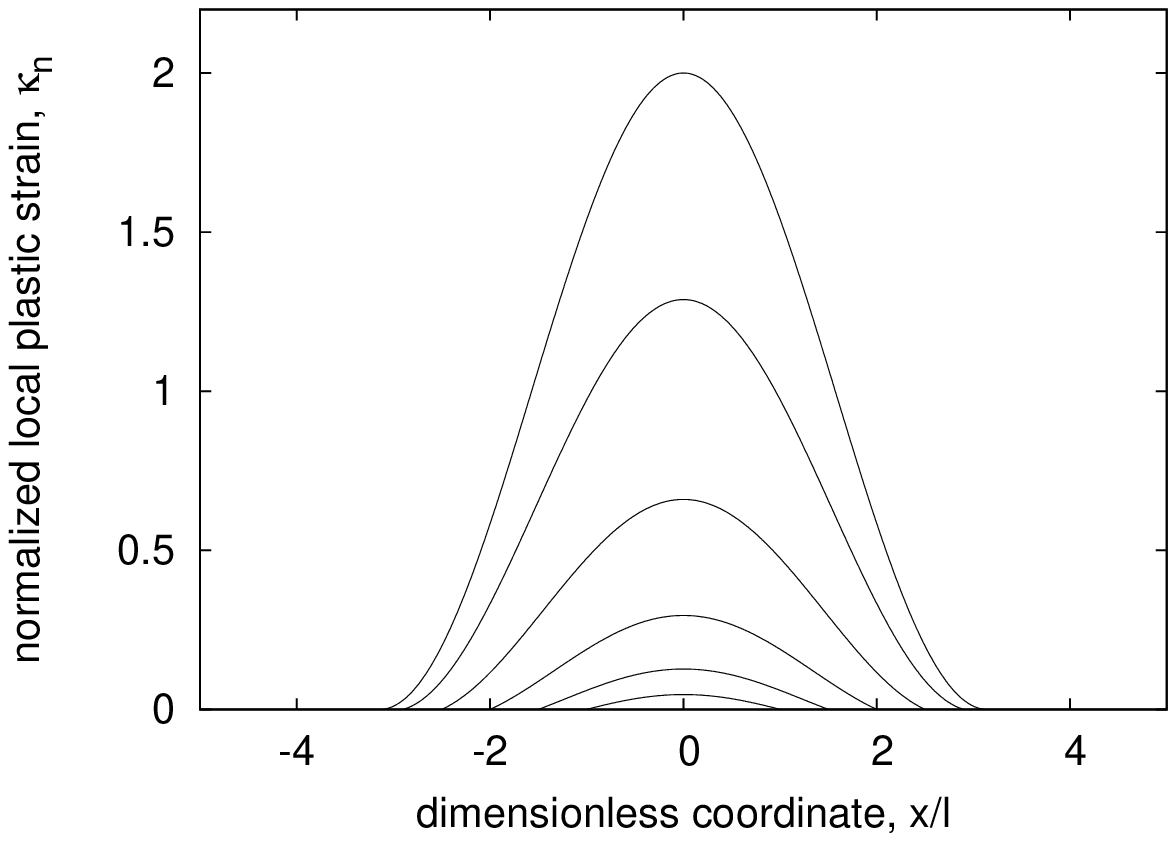}
&
\includegraphics[scale=0.7]{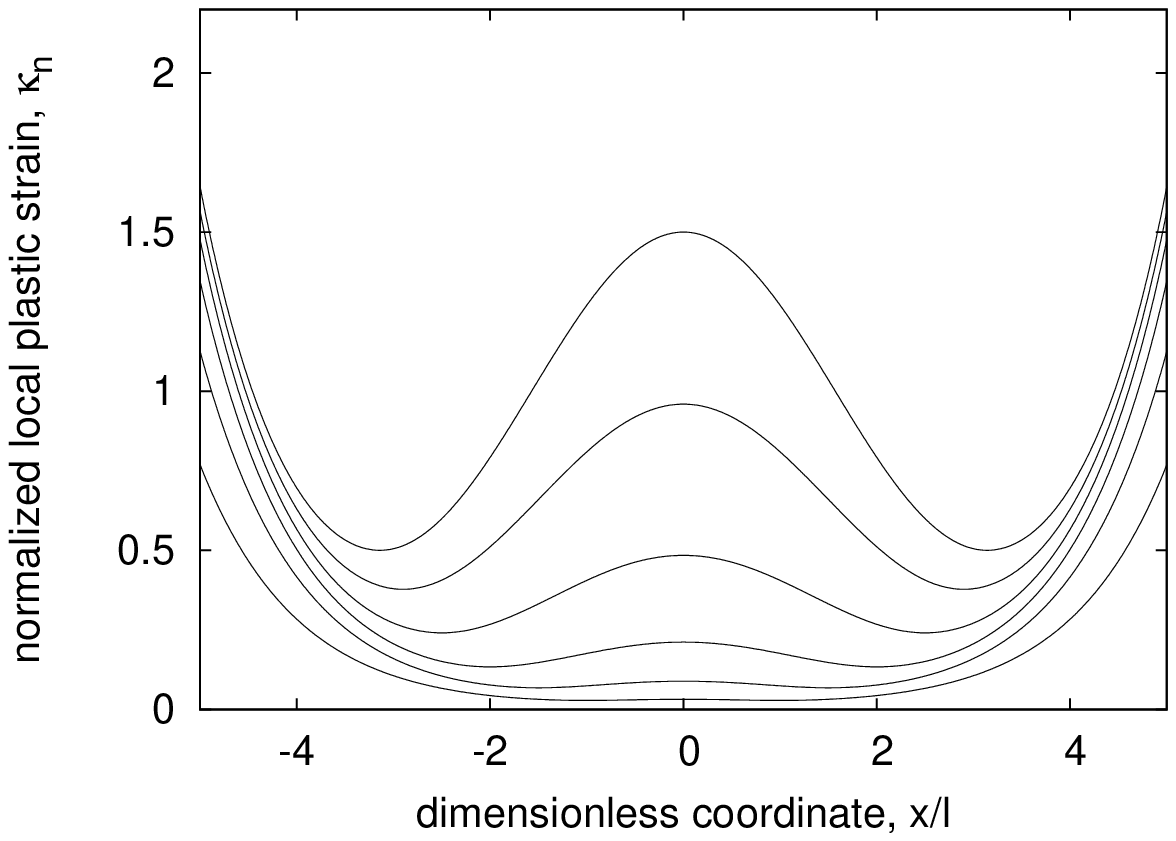}
\end{tabular}
\caption{Modified implicit gradient model, quadratic stress distribution: Evolution of normalized plastic strain profile for $m=2$ and $\lamg=5$ --- (a) local plastic strain, (b) nonlocal plastic strain}
\label{fig:ImpA2-kap}
\end{figure}

\begin{figure}
\centering
\begin{tabular}{cc}
(a) & (b) 
\\
\includegraphics[scale=0.7]{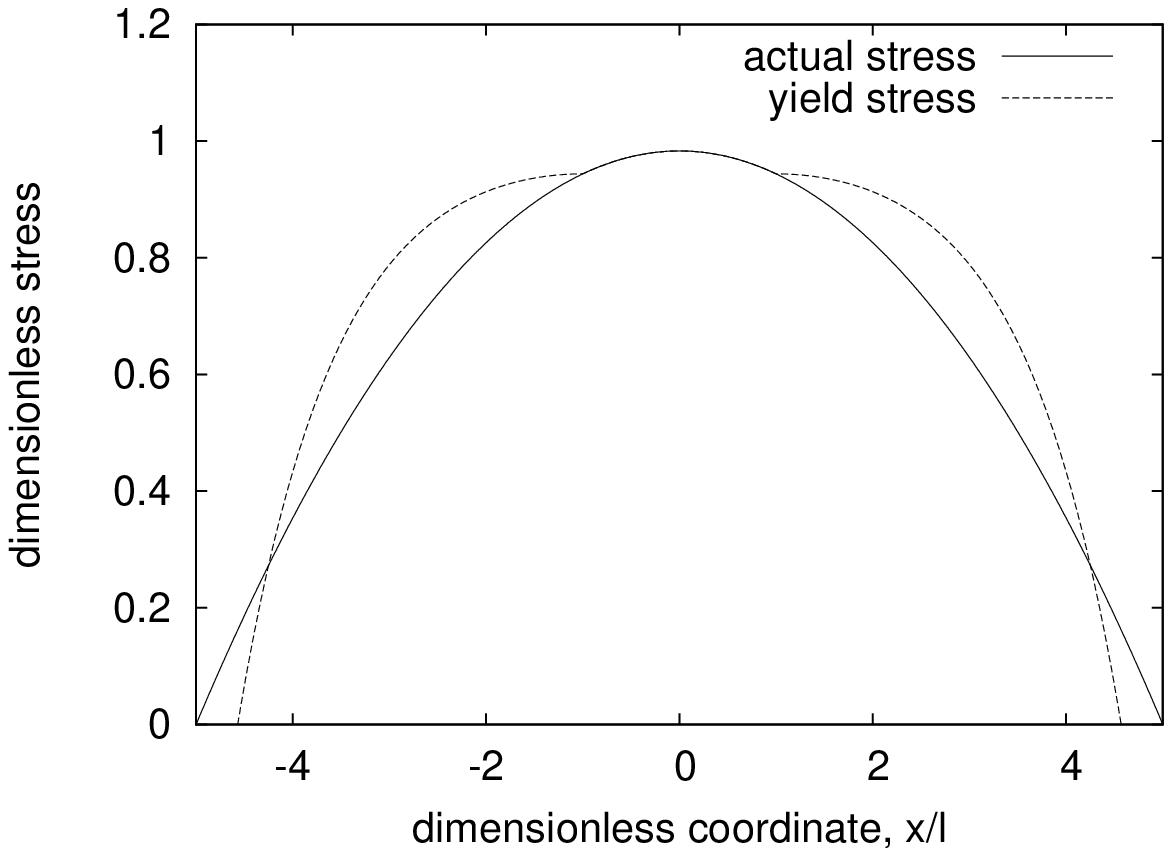}
&
\includegraphics[scale=0.7]{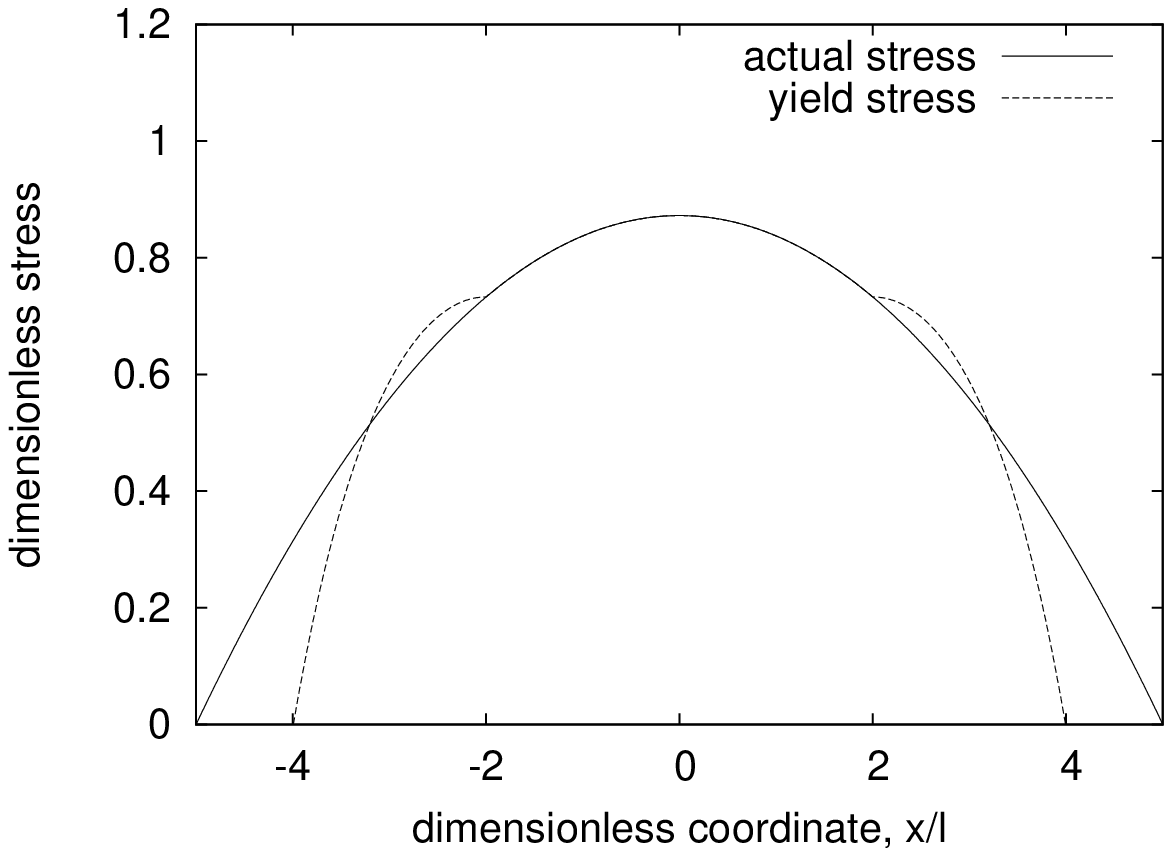}
\end{tabular}
\caption{Modified implicit gradient model, quadratic stress distribution: Distribution of normalized actual stress $\sig/\sig_0$ and yield stress $\sig_Y/\sig_0$ at
plastic zone size (a) $\lamp=1$, (b) $\lamp=2$}
\label{fig:ImpA-ss2}%
\end{figure}

The evolution of the local and nonlocal plastic strain profiles is depicted in Fig.~\ref{fig:ImpA2-kap}
for the specific parameter values $m=2$ and $\lamg=5$. Similar to all previously discussed models,
the plastic zone expands and the local plastic strain grows monotonically. However, a deeper examination
of the solution reveals that the current model
exhibits a pathological behavior, which becomes apparent if we plot the distribution of the actual stress
and the current yield stress (Fig.~\ref{fig:ImpA-ss2}). In the plastic zone $\Ip$ that extends from
$\xi=-\lamp$ to $\xi=\lamp$, the actual stress $\sigma$ is equal to the yield stress $\sigma_Y$,
which is quite natural because the yield condition $\sig=\sig_Y$
is in fact the differential equation from which the solution has been
calculated. In the elastic zone, the solution should comply with the condition of plastic admissibility,
requiring that the actual stress must not exceed the yield stress. 
As seen in Fig.~\ref{fig:ImpA-ss2}, 
the inequality $\sigma\le\sigma_Y$ is satisfied in the proximity of the
plastic zone but violated farther away. The problem is caused by the modified ``boundary'' condition, which enforces
a vanishing derivative of nonlocal plastic strain 
at the boundary of the plastic zone. 
The nonlocal plastic strain has then local minima at the points separating the plastic zone from the elastic ones
(Fig.~\ref{fig:ImpA2-kap}b), 
and its value in the elastic zones, calculated from (\ref{eq:igp4}),
increases with increasing distance from the plastic zone (note the hyperbolic cosine function in
the second line of (\ref{eq103})). The local plastic strain in the elastic zones vanishes and the nonlocal
plastic strain becomes large, which results into a dramatic reduction of the yield stress.
So even though the actual stress in the 
elastic zone is relatively low,  the plastic admissibility condition is violated and additional plastic
zones would be formed. This is of course a non-physical, unacceptable artefact, and thus the model with
the modified boundary condition cannot be considered as a viable alternative to the standard implicit
gradient formulation.

\subsection{Piecewise linear stress distribution}

For completeness, the results obtained with the modified implicit gradient model for a piecewise linear
distribution of stress are presented:
\bea
\delta_1 &=& \lamp-\frac{m}{\mu}\tan\frac{\lamp}{2\mu}
\\
\kaptb(\xi) &=&
\left\{ \begin{array}{ll} 
\displaystyle\frac{\phi}{\lamg}\left(\vert\xi\vert-\delta_1+\mu\tan\frac{\lamp}{2\mu}\cos\frac{\xi}{\mu}-\mu\sin\frac{\vert\xi\vert}{\mu}\right) & \mbox{ for } \xi\in\Ip=(-\lamp,\lamp)
\\
\displaystyle\frac{\phi}{\mu\lamg}\tan\frac{\lamp}{2\mu}\cosh(\vert\xi\vert-\lamp) & \mbox{ for } \xi\in\Ie={\cal L}\setminus\Ip
\end{array}\right.
\\[3mm]
\kapt(\xi) &=&
\left\{ \begin{array}{ll} 
\displaystyle\frac{\phi}{\lamg}\left(\vert\xi\vert-\delta_1+\frac{m}{\mu}\tan\frac{\lamp}{2\mu}\cos\frac{\xi}{\mu}-\frac{m}{\mu}\sin\frac{\vert\xi\vert}{\mu}\right)
 & \mbox{ for } \xi\in\Ip=(-\lamp,\lamp)
\\
0 & \mbox{ for } \xi\in\Ie={\cal L}\setminus\Ip
\end{array}\right.
\\
2\lambda_{p,\max{}}&=&2\pi\mu
\\
\frac{u_p}{l\kap_f}&=&\frac{\phi\lamp}{\lamg}
(\lamp-2\delta_1)
\eea
Since the model exhibits a pathological behavior, similar to the preceding subsection,
the graphical presentation and discussion of the results can be omitted.

\section{Summary and conclusions}

In this paper, analytical solutions have been derived for localization
of plastic strain described by gradient plasticity models in one spatial
dimension. 
In previous studies, this problem was usually investigated for the simplest
case of a bar with uniform properties, and localization was treated
as a bifurcation from the uniform solution. Here we have considered 
a nonuniform distribution of stress along the bar due to a variable
cross-sectional area. To keep the problem tractable, the
stress distribution has been assumed to be quadratic or piecewise linear.
The first case represents a smooth variation of the sectional area and
the second case   represents a non-smooth but still differentiable variation.
Alternatively, the resulting mathematical problem can be interpreted
as the description of a bending beam subjected to a uniform or concentrated
lateral load. The function specifying the stress or bending moment distribution 
contains a parameter with the dimension of length, which reflects the
length scale of the structural geometry (e.g., the span of the beam in
the case of bending).

Both explicit and implicit formulations of gradient plasticity have been
investigated, with a second-order or fourth-order enhancement in the
explicit case and with different types of boundary conditions in the
implicit case. The gradient term incorporated into the softening law
contains a new material parameter, which reflects the intrinsic material
length scale. 

The results obtained with the second-order and fourth-order explicit
model, as well as with the  implicit model that applies homogeneous Neumann
boundary conditions at the physical boundary of the body of interest,
are qualitatively similar. Plastic yielding starts as soon as the yield stress
is attained at the weakest section, the plastic zone continuously
grows from that section and expands up to a maximum size that directly
depends on the intrinsic material length and corresponds to the solution
of the bifurcation problem for an idealized bar with perfectly uniform
properties. At early stages of the inelastic process, the growth of the
plastic zone occurs at increasing axial force (and thus at increasing stress).
Even though the local material response is postulated as softening, the global response
at the structural level is initially hardening, and only after a certain 
critical size of the plastic zone has developed, the global response turns
into softening. The maximum level of the axial force thus 
depends not only on the yield stress and on the area of the weakest section
but also on the distribution of the area in the vicinity of that section,
which is  in perfect agreement with the concept of nonlocal interactions
in the material microstructure,
described by the gradient terms. The increase of the ultimate load
as compared to the local model (which would fail abruptly with plastic yielding
fully localized into the weakest section) directly depends on the ratio
between the length parameters characterizing the material and the geometry.
 If this ratio tends to zero, the ultimate load tends to the ``locally''
determined one. 

It is thus concluded that the behavior of the models mentioned in the
previous paragraph is perfectly reasonable and all of them can describe
a gradual development of the plastic zone, at the structural
level accompanied by a transition
from hardening to softening. In contrast to that, a modified version of the
implicit gradient model with homogeneous Neumann conditions prescribed at the
boundary of the plastic zone (and not at the physical boundary of the body)
exhibits a pathological behavior related to the unrealistic distribution
of nonlocal plastic strain in the elastic zone, and it would lead to spurious
yielding at points far from the ``main'' plastic zone, which are under low stress
but their yield stress is artificially reduced. Therefore, this modified formulation
of implicit gradient plasticity is not acceptable.   

In the future, the localization problem will be reformulated using a variational approach,
which permits a more systematic treatment of discontinuities and reduces
the regularity requirements. This is important e.g.\ 
for a rigorous justification
of the admissibility conditions for the fourth-order gradient model,
which are in this paper postulated based on intuition.

\subsection*{Acknowledgment}
Financial support of
the Ministry of Education of the Czech Republic under the Research Plan 
MSM 6840770003 is gratefully acknowledged.

\providecommand{\bibhead}[1]{}

\end{document}